\documentclass[pre,twocolumn, amsmath,amssymb,superscriptaddress]{revtex4}
\usepackage{graphicx}
\usepackage{amssymb}
\usepackage{epstopdf}
\usepackage{makecell}
\usepackage{bbold}
\usepackage{soul}
\usepackage{bm}
\usepackage[dvipsnames]{xcolor}
\DeclareMathAlphabet\mathbfcal{OMS}{cmsy}{b}{n}
\usepackage{accents}

\begin{document}
\title{Generic theory of interacting, spinning, active polar particles: a model for cell aggregates}
\author{Quentin Vagne}
\affiliation{University of Geneva, Quai Ernest Ansermet 30, 1205 Gen\`eve, Switzerland}
\author{Guillaume Salbreux}
\affiliation{University of Geneva, Quai Ernest Ansermet 30, 1205 Gen\`eve, Switzerland}
\date{\today}

\begin{abstract}
We present a generic framework for describing interacting, spinning, active polar particles, aimed at modelling dense cell aggregates, where cells are treated as polar, rotating objects that interact mechanically with one another and their surrounding environment. Using principles from non-equilibrium thermodynamics, we derive constitutive equations for interaction forces, torques, and polarity dynamics. We subsequently use this framework to analyse the spontaneous motion of cell doublets, uncovering a rich phase diagram of collective behaviours, including steady rotation driven by flow-polarity coupling or interactions between polarity and cell position.
\end{abstract}
\maketitle
\section{Introduction}
Cells inside tissues act as motile organising elements. They exert forces on one another and on surrounding substrates, allowing for the emergence of organised collective migration that is necessary for embryo morphogenesis and tissue regeneration \cite{collective_mig}.  Cell polarity is a fundamental aspect of such tissue self organisation. For instance, planar polarity in two dimensional tissues acts as a global coordinator of morphogenesis \cite{zallen_planar_2007}, while apico-basal polarity is essential to regulate the form and the function of epithelial tissues \cite{buckley_apicalbasal_2022}. Cell polarity is also directly involved in setting the direction of cell migration at the individual \cite{doi:10.1126/science.1092053} and collective \cite{jain_role_2020} scale.

There are multiple theoretical frameworks that can be used to describe collective cell motion, from continuous approaches to agent-based models \cite{active_matter}. At the continuous level, generic active gels theories have been developed \cite{julicher_hydrodynamic_2018}. While they can be successfully used to study the physics of tissues with large number of cells \cite{etournay2015interplay, morita2017physical, bowick2022symmetry}, they are not adequate to describe small aggregates, such as in the early stages of embryogenesis or during organoid development, or systems with highly inhomogeneous organisation. In such cases, cells need to be described as individual entities. In the active Brownian particles (ABP) framework \cite{active_part,smeets_cil_2016}, cells are seen as self-propelled point-like objects, constituting dilute systems where cells interact through transient collision, hydrodynamic interactions and/or chemotactic behaviour \cite{saha_pairing_2019} or denser systems interacting through an effective passive potential \cite{henkes2011active, alert2020physical}. When the passive potential includes polarity, interesting self-organising properties appear to emerge \cite{germann_ya_2019,nissen_theoretical_2018}. 
\begin{figure}
\includegraphics[width=\linewidth]{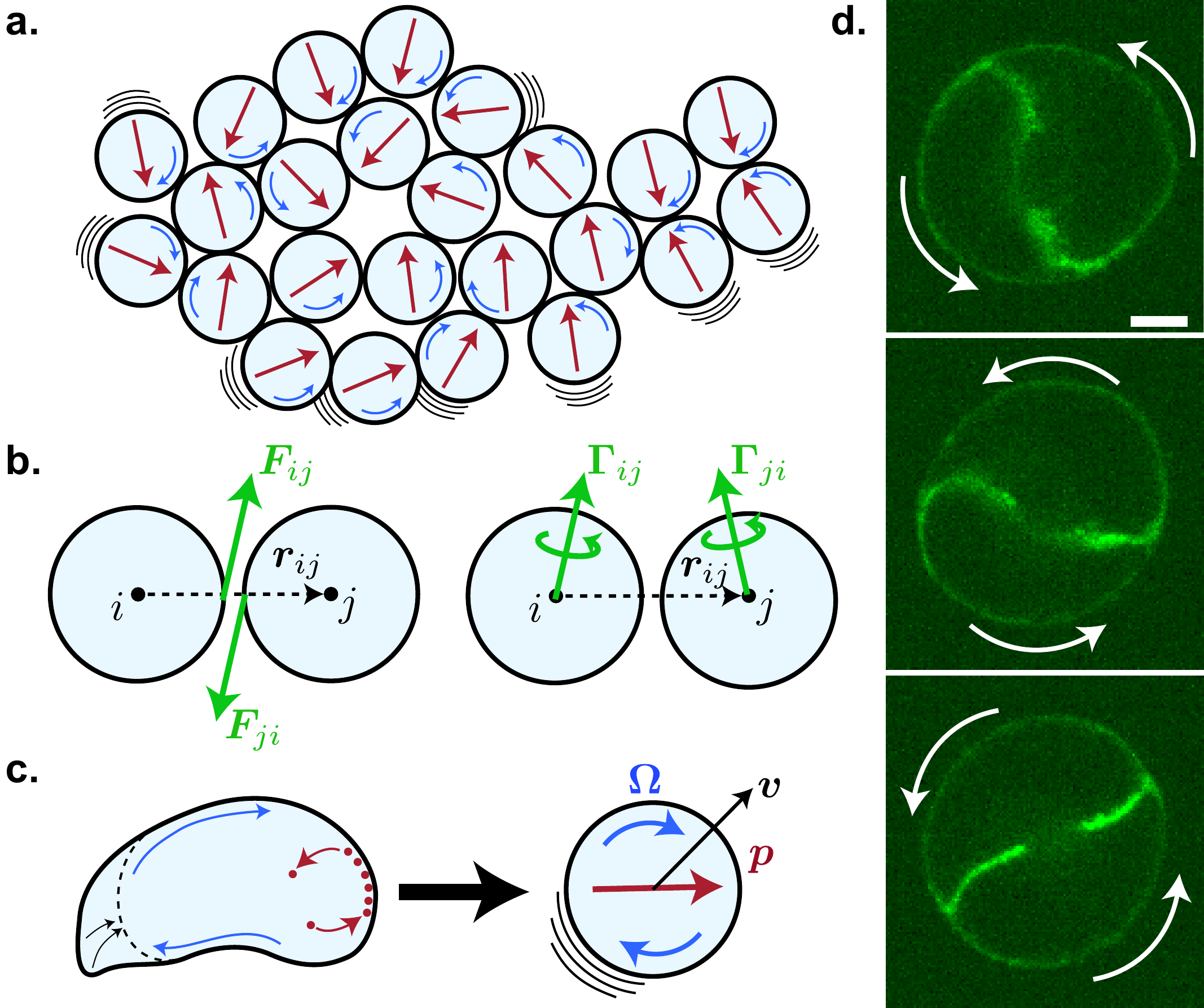}
\caption{\textbf{a.} Scheme of a biological tissue, approximated as a collection of interacting particles. Red arrows indicate polarity, blue arrows indicate rotation. \textbf{b.} A particle $i$, in contact with particle $j$, is subjected to a force $\bm F_{ij}$ and a torque $\bm \Gamma_{ij}$. \textbf{c.} Schematic of possible cellular processes giving rise to polarity $\bm p$ (red arrow), angular velocity $\bm \Omega$ (blue arrows) and center of mass velocity $\bm v$ (black arrow). In the cell schematics on the left, black arrows: cell membrane deformation, blue arrows: internal flows, red dots: molecules establishing a cell polarity axis, red arrow: protein turnover.  \textbf{d.} Snapshots of a doublet of MDCK cells, spontaneously rotating in Matrigel. 20min between frames, E-cadherin labelled in green (courtesy D. Riveline). scale bar: 5 $\mu$m. }\label{fig1}
\end{figure}
Generic active effects that arise in dense ensembles of active polar particles, describing cells which adhere and are constantly in contact, remain however to be explored (Fig.\ref{fig1}a).

In addition, spontaneous rotational motion is often observed in assemblies of cells, in whole tissues \cite{cetera_epithelial_2014,founounou_tissue_2021} or in controlled in-vitro settings with either cell aggregates \cite{lo_vecchio_spontaneous_2024, guillamat_integer_2022,fernandez_surface-tension-induced_2021} or doublets of cells in two \cite{brangwynne_symmetry_2000,lee_neighbor-enhanced_2021} or three \cite{lu_polarity-driven_2022} dimensions. In cultures of human mammary epithelial cells, global rotational motion is necessary for the proper deposition of extracellular matrix \cite{wang_ecmandrotation_2013}. In zebrafish embryos, rotation of cell pairs occurs in the lateral line \cite{erzberger_mechanochemical_2020,Kozak_2023}. In all of these cases, an understanding of the translational and rotational motion of individual cells is required to correctly capture the physics of the entire aggregate.

These observations raise the question of the collective dynamics which emerges from the interaction of physical objects which possess a polarity and a spinning degree of freedom, and interact out-of-equilibrium. To address this question, we propose here generic equations for the ``interacting active particles'' framework where individual objects are described as discrete, active, polarised, and spinning particles which can both translate and rotate and which directly exchange forces, torques and have coupled polarity dynamics (Fig. \ref{fig1}a, c). We use linear irreversible thermodynamics to derive generic constitutive relations for active and passive interaction forces, torques (Fig.\ref{fig1}b) and polarity dynamics of an ensemble of particles, intended to describe a cell aggregate. We obtain terms that depend, at the lowest order, on the relative positions, polarity of the particles and molecule numbers, and which respect the symmetries of the system. 
Here we use our framework to study the simplest case of the motion of a pair of adhering cells, exhibiting a rotational movement  \cite{lu_polarity-driven_2022} (Fig. \ref{fig1}d).

\section{Generic description of $N$ active particles}
\subsection{Equations of motions, free energy and equilibrium quantities}
We choose a coarse-grained description where each object $i$ is characterised by a particle with velocity $\bm v_i$, angular velocity $\bm \Omega_i$, polarity $\bm p_i$ (Fig.\ref{fig1}c), mass $m_i$, and spin angular momentum $\bm L_i$. We describe a system of $i=1...N$ interacting particles. The force and torque balance equations are: 
\begin{align}
 \bm F_i=&\bm F^{f}_i+\sum_{j\in \mathcal{N}_i}\bm F_{ij} =m_i \frac{d\bm v_i}{dt} ,~\label{force_bal} \\
\bm \Gamma_i=&\bm \Gamma^f_i+\sum_{j\in \mathcal{N}_i}\bm \Gamma_{ij}=\frac{d \bm L_i}{dt}~,\label{torque_bal}
\end{align}
where $\mathcal{N}_i$ is the set of particles in contact with particle $i$, $\bm F_i$ and $\bm \Gamma_i$ are the total force and total torque relative to the particle center of mass acting on particle $i$, $\bm F_i^f$, $\bm \Gamma_i^f$ are the force and torque exerted by the surrounding fluid on particle $i$, and $\bm F_{ij}$, $\bm \Gamma_{ij}$ are the force and torque exerted by particle $j$ on particle $i$ which satisfy (see Appendix \ref{micro-macro-appendix}, Eqs. \ref{3rdlawforce},\ref{3rdlawtorque}):
\begin{align}
\mathbf{F}_{ij}+\mathbf{F}_{ji}=0,~ \boldsymbol{\Gamma}_{ij}+\boldsymbol{\Gamma}_{ji}=\mathbf{r}_{ij}\times \mathbf{F}_{ij}~.
\end{align}
In the following, we assume that the fluid acts as a momentum sink without an internal dynamics. Further assuming that the free energy for $N$ particles only involves pair-wise interactions, we write:
\begin{multline}
f= \sum_{i=1}^{N}\left( \frac{1}{2}m_i\bm v_i^2+ \frac{1}{2}\bm L_{i}\cdot \bm \Omega_i \right)+\sum_{i=1}^{N} f_0\left( \bm p_i, \{N^{(i)}_k\} \right)\\
 + \sum_{\langle i, j \rangle}f_{\text{int}}(\bm p_i,\bm p_j,\bm r_{ij},\{N^{(i)}_k\},\{N^{(j)}_k\})~, \label{free_energy}
\end{multline}
which is the sum of the kinetic energy of translation and rotation of individual particles of mass $m_i$, a free energy $f_0$ per particle, and an interaction energy $f_{\text{int}}$ between pairs $\langle i,j \rangle$ in contact. We introduced quantities $N^{(i)}_k$, the number of molecules of type $k\in \{1,2,..,n\}$, present inside object $i$, involved in a single chemical reaction that drives the system out of equilibrium. The interaction energy $f_{\text{int}}$ satifies the symmetry property $f_{\text{int}}(\bm p_i,\bm p_j,\bm r_{ij},\{N^{(i)}_k\},\{N^{(j)}_k\})=f_{\text{int}}(\bm p_j,\bm p_i,\bm r_{ji},\{N^{(j)}_k\},\{N^{(i)}_k\})$.
We define the chemical potential difference $\Delta \mu_i$ associated to the chemical reaction in cell $i$ as:
\begin{equation}
\begin{aligned}
\Delta\mu_i & = \sum_{k}\nu_k \left(\Delta\mu_{ik}^{(0)} +\sum_{j \neq i}\Delta \mu^{\langle ij \rangle}_{ik}\right)\text{ with:}\\
\Delta\mu_{ik}^{(0)} & =\frac{\partial f_0}{\partial N_k^{(i)}}~,\\
\Delta \mu^{\langle ij \rangle}_{lm} & = \frac{\partial } {\partial N_{m}^{(l)}}f_{\text{int}}(\bm p_i,\bm p_j, \bm r_{ij},\{N^{(i)}_k\},\{N^{(j)}_k\})~,
\end{aligned}
\end{equation}
where $\nu_k$ are the stoichiometric coefficients of the reaction. Similarly, we define the general force $\bm h_i$ acting on polarity $\bm p_i$:
\begin{multline}
\bm h_i  =\bm h_i^{(0)}+\sum_{j\neq i}\bm h_i^{\langle ij \rangle}\text{ with }\bm h^{(0)}_i  = -\frac{\partial f_0}{\partial \bm p_i}\\
\text{and }\bm h^{\langle ij \rangle}_{k}  =-\frac{\partial }{\partial \bm p_k}  f_{\text{int}}(\bm p_i,\bm p_j,\bm r_{ij},\{N^{(i)}_k\},\{N^{(j)}_k\})~.\label{hdef}
\end{multline}
Equilibrium interaction forces and torques $\bm F_{ij}^{\rm eq}$ and $\bm \Gamma_{ij}^{\rm eq}$, can also be obtained from the free energy $f$ (Appendix \ref{eq_fo_to}, Eq.\ref{eq_tor_for}):
\begin{equation}
\bm F_{ij}^{eq}  = \frac{\partial f_{\text{int}}}{\partial \bm r_{ij}}~,~\bm \Gamma_{ij}^{eq} = -\bm h^{\langle ij\rangle}_i \times \bm p_i~.\label{eq_for_tor_main}
\end{equation}
\subsection{Fluxes and forces out of equilibrium}
We now obtain the time derivative of the free energy $df/dt$, in order to identify the generalised fluxes and forces driving the system out of equilibrium. The time derivative of the kinetic energy terms is given by (Appendix \ref{kinetic-section}):
\begin{equation}
\frac{d}{dt}\left(\frac{1}{2}m_i\bm v_i^2\right) = \bm F_i \cdot \bm v_i~,~\frac{d}{dt}\left( \frac{1}{2}\bm L_i \cdot \bm\Omega_i\right) = \bm \Gamma_i \cdot \bm \Omega_i  ~.\label{kinetic}
\end{equation}
The numbers of molecules $N_k^{(i)}$ and particle polarity $\bm p_i$ vary according to:
\begin{align}
\frac{dN_{k}^{(i)}}{dt}& =\nu_k r_i~,\label{reaction}\\
\frac{d\bm p_i}{dt} & = \bm \Omega_i \times \bm p_i + \frac{D\bm p_i}{Dt}~.\label{corot}
\end{align}
Here, $r_i$ is the reaction rate inside particle $i$. The time evolution of the polarity $\bm p_i$ includes the term $D\bm p_i/Dt$ which represents any polarity change that differs from simply following the rotation of the particle. In the case of a cell, such intrinsic polarity remodelling does not necessarily require material flows inside the cell, but can arise from reaction-diffusion based mechanisms like depicted in Fig.\ref{fig1}c. Plugging in the force and torque balance equations (Eqs.\ref{force_bal}-\ref{torque_bal}), the equilibrium forces and torques (Eq. \ref{eq_for_tor_main}), the kinetic terms (Eq. \ref{kinetic}), and the terms from Eqs. \ref{reaction}, \ref{corot}, we find that $df/dt$ can be put into the following form:
\begin{equation}
\frac{df}{dt} = \sigma_c + \sigma_{f} +\sigma_{int}~, \label{entropy}
\end{equation}
where we have introduced the entropy productions $-\sigma_c$ (chemical reaction and polarity dynamics), $-\sigma_f$ (mechanical interactions with the fluid), and $-\sigma_{int}$ (mechanical interactions between particles),  given by:
\begin{equation}
\begin{aligned}
\sigma_c& = \sum_i \left( \Delta\mu_i r_i -\bm h_i \cdot \frac{D\bm p_i}{Dt}\right)~,\\
\sigma_f & = \sum_{i} \left(  \bm F^{f}_i\cdot \bm v_i + \bm \Gamma^{f}_i \cdot \bm \Omega_i\right)~, \\
\sigma_{\rm int} & =  \sum_{\langle i, j\rangle}\left( \bm F^d_{ij}\cdot \Delta \bm v_{ij} + \bm \Gamma_{ij}^d\cdot(\bm \Omega_i-\bm \Omega_j) \right)~.
\end{aligned}\label{entropy_terms}
\end{equation}
We have also introduced a sliding velocity, as well as deviatoric forces and torques:
\begin{equation}
\begin{aligned}
\Delta \bm v_{ij} & = \bm v_i-\bm v_j+\frac{1}{2}(\bm \Omega_i+\bm \Omega_j)\times \bm r_{ij}~,\\
\bm F_{ij}^d & = \bm F_{ij} - \bm F^{eq}_{ij}~, \\
\bm \Gamma_{ij}^d & =\bm \Gamma^a_{ij} + \frac{1}{2}(\bm h^{\langle i j \rangle}_i\times \bm p_i-\bm h^{\langle i j \rangle}_j\times \bm p_j)~.
\end{aligned}\label{deviatoric}
\end{equation}
$\Delta \bm v_{ij}$ represents an effective difference of velocities between particles $i$ and $j$, that occurs due to rotation and/or translation. It is coupled to the out of equilibrium (also called deviatoric) interaction force $\bm F_{ij}^d$ between the cells, which is the part of the total interaction force $\bm F_{ij}$ that is non zero only when the system is out of equilibrium. $\bm \Gamma^a_{ij}$ is the antisymmetric part of the torque $\bm \Gamma_{ij}$ (Eq.\ref{gamma_a}, the symmetric part being $(\bm r_{ij}\times \bm F_{ij})/2$). $\bm \Gamma_{ij}^d$ is the deviatoric part of $\bm \Gamma_{ij}^a$ and is coupled with $\bm \Omega_i-\bm \Omega_j$ which represents the particles rolling on one another. Both $\Delta v_{ij}$ and $\Omega_i-\Omega_j$ are invariant by solid translation and rotation of space. Using Eqs.\ref{gamma_a},\ref{eq_tor_for},\ref{deviatoric} the total torque is related to the dissipative torque through:
\begin{align}
\bm\Gamma_{ij}=\bm \Gamma_{ij}^{eq}+\bm \Gamma_{ij}^d+\frac{1}{2}\bm r_{ij}\times\bm F_{ij}^d~.
\end{align}
The list of fluxes and forces driving the system out of equilibrium can be identified from Eq.\ref{entropy_terms}, and are the following ones:
\begin{equation}
\begin{array}{!{\vrule width 1pt} c | c !{\vrule width 1pt}}
\Xhline{1pt}
 \text{Force} & \text{Flux}  \\
 \Xhline{1pt}
 \Delta \mu_i & r_i  \\
  \hline
 -\bm h _i & D \bm p_i/Dt \\
 \Xhline{1pt}
  \bm \Omega_i & \bm \Gamma^f_i  \\
 \hline
 \bm v_i & \bm F^f_i   \\
 \Xhline{1pt}
 \Delta \bm v_{ij} & \bm F_{ij}^d  \\
 \hline
 \bm \Omega_i - \bm \Omega_j & \bm \Gamma_{ij}^d \\
 \Xhline{1pt}
\end{array} \label{signatures}
\end{equation}
We arranged the fluxes intro three groups, which are:
\begin{itemize}
\item ($r_i$, $D\bm p_i/Dt$): Chemical reactions and polarity dynamics, internal to individual particles.
\item ($\bm F_i^f$, $\bm \Gamma_i^f$): Mechanical interactions between particles and the surrounding fluid. As the fluid is considered to be at rest, these forces and torques are not invariant under rotation and translation of the particles. The corresponding generalized forces $\bm\Omega_i$ and $\bm v_i$ are angular velocities and velocities relative to the fluid.
\item ($\bm F_{ij}^d$, $\bm \Gamma_{ij}^d$): Direct, pair-wise, mechanical interactions between particles, invariant by translation and rotation of cell pairs.
\end{itemize}
\subsection{Constitutive relations}
\label{consteq_section_main}
The dynamics of the system is given by a set of constitutive equations, which gives the expression of the fluxes when the system is out of equilibrium. Assuming that the system is only weakly out of equilibrium, one can write the generalised fluxes $J_k$ as linear combinations of the generalised forces $F_k$ \cite{de2013non, julicher_hydrodynamic_2018}:
\begin{equation}
J_k = O_{kl}F_l~,\label{eqlinearresponse}
\end{equation}
where $J_k$/$F_k$ is either a scalar flux/force or the coordinate $\alpha$ of a vectorial flux/force, and $O_{kl}$ are the coupling coefficients. Thus, a coupling corresponds to a triplet $\{J_k, F_l, O_{kl}\}$. The coupling coefficients verify the Onsager reciprocity relations:
\begin{equation}
O_{kl} = \epsilon_k \epsilon_l O_{lk}~,\label{onsager_symmetry}
\end{equation} 
where $\epsilon_k$ is the time signature of flux $k$ shown on Table \ref{signatures_detailed}.  For a system of $N$ beads, in principle every flux in Table \ref{signatures} could be coupled to every force. In order to ensure that the theory is tractable, and since we wish to describe dense aggregates where particles are in direct contact with one another, we assume that only particles in contact interact, and
we introduce an out-of-equilibrium pair-wise interaction condition, which we define as such: each coupling triplet $\{J_k, F_l, O_{kl}\}$ can only depend on a pair of interacting particles $i,j$, or on a particle $i$ and the fluid $f$. In addition, we assume that coupling terms between fluxes intrinsic to particle $i$ and forces intrinsic to particle $i$ (here pairs  $r_i$, $\Delta\mu_i$ and  $D\bm p_i/Dt$, $\bm h_i$) only depend on state variables of that particle.
Altogether, this allows to determine which coefficients $O_{kl}$ are non-zero (see Appendix \ref{consteq_appendix}). The form of the resulting constitutive equations is given in Eq. \ref{coupling_array}. Since the Onsager theory is built as an extension around an equilibrium point, the coefficients may depend on equilibrium quantities, which are the $\bm r_{ij}$ and $\bm p_i$ vectors, as well as molecules numbers $N_{k}^{(i)}$. We assume that coupling coefficients associated to a particle $i$ only depend on the properties of particle $i$ or its direct neighbours, and coupling coefficients associated to the pair $\langle i,j\rangle$ only depend on the properties of particles $i$ and $j$. We obtain the expressions of the coefficients (Eq. \ref{coeffs}) at lowest order in $\bm r_{ij}$, $\bm p_i$, $N_{k}^{(i)}$; while these terms are not necessarily small, this provides with a systematic procedure to introduce couplings of increasing complexity. We ultimately obtain the extended constitutive equations, which for the sake of readability are shown in Appendix \ref{consteq_appendix} (Eq.\ref{constitutive_N_full}). Together with Eqs.\ref{force_bal}-\ref{torque_bal}, Eq.\ref{corot}, and expressions for $f_0$ and $f_{\text{int}}$, these constitutive equations fully determine the time evolution of a system of $N$ particles.

In the following, for the sake of simplicity, we neglect reactive couplings with the molecular field that introduce non-linearities in the constitutive equations ($\chi$, $\gamma$, $\gamma'$, $\gamma''$), the polarity dependent cross-coupling dissipative terms $\beta$, $\beta'$, and the dissipative cross-coupling coefficient $\alpha'$. 
We
assume that all particles have the same constant chemical potential difference, such that $\Delta\mu_i=\Delta\mu$. As a result, some active terms in the interaction forces and torques vanish, as they depend on the differences of chemical potential $\Delta\mu_i-\Delta\mu_j$ between interacting particles. 
The reaction rate $r_i$ must be controlled for $\Delta\mu$ to remain constant, so we do not consider the first line of Eq.\ref{constitutive_N_full}. Since we will consider applications where the norm of the polarity $\bm p_i$ and of the distance $\bm r_{ij}$ are fixed (by introducing Lagrange multipliers in the expressions of $f_0$ and $f_{\text{int}}$), we do not further consider terms in $\lambda'_r$, $\zeta$. This leads to the following simplified set of constitutive equations:
\begin{equation}
\begin{aligned}
\bm F_i^f  =& \Delta\mu \lambda \bm p_i -\xi \bm v_i -\eta \bm h_i, \\
\bm \Gamma_i^f  = & -\xi_r \bm \Omega_i, \\
\bm F_{ij}^d  = & \Delta\mu\left[\lambda'(\bm p_i-\bm p_j)\right]-\xi_{||}\Delta \bm v_{ij}+\eta' (\bm h_i -\bm h_j), \\
\bm \Gamma_{ij}^d  = &\Delta\mu\left[\mu_r\bm r_{ij}\times(\bm p_i+\bm p_j)+\mu'_r\bm p_i\times\bm p_j\right]\\
&-\xi'_r(\bm \Omega_i - \bm \Omega_j),\\
\frac{D\bm p_i}{Dt}  = & \sum_{j \in \mathcal{N}_i}\left(\Delta\mu\left[\zeta' \bm p_j + \zeta_r \bm r_{ij}\right] + \eta'\Delta \bm v_{ij}\right) -\eta \bm v_i +\alpha \bm h_i.
\end{aligned}\label{constitutive_N}
\end{equation} 
The term in $\lambda$ is a self-propelling force against the external medium usually considered in ABPs models \cite{alert2020physical}. Here we also have introduced an active interaction force (coupling in $\lambda'$) and active interaction torques (couplings in $\mu_r$, $\mu'_r$). They correspond to active mechanical interactions between pairs of particles. They may originate from the complex patterns of microscopic forces that objects exert on one another. In the case of epithelial cells, for instance, forces are transmitted through adherens junctions coupled to cortical flows \cite{Noordstra2023,Arslan2024}. In colonies of Myxococcus xanthus bacteria, individual bacteria use protrusions called pili to pull on their neighbours \cite{zhang2012}.
 In Eq.\ref{constitutive_N} we find two active polarity couplings which couple the dynamics of a particle polarity to the polarity of its neighbours (coupling $\zeta'$), and to the vector joining a particle to its neighbours (coupling $\zeta_r$). Such terms may describe reorientation of cell polarity as a response to contact with other cells \cite{cote2022}. $\xi$, $\xi_r$ are dissipative coefficients corresponding to translational and rotational friction against the external medium. $\xi_{||}$ and $\xi'_r$ are dissipative coefficients describing relative friction between the particles. $\alpha$ is a rotational viscosity for the polarity. $\eta$,  $\eta'$ are flow-polarity couplings between the velocity and polarity. $\eta$ is called a velocity-alignment term in models of ABPs \cite{PhysRevE.74.061908, alert2020physical}. The term in $\eta'$ represents a polarity flow coupling due to relative sliding between neighbouring particles. In addition to active self-propulsion, it can capture the tendency of particles to align with the velocity of their neighbours  \cite{Nestor_2013_neighbour_align,bruckner_learning_2021}. In the following, the chemical potential difference $\Delta\mu$ is absorbed implicitly in the definition of the active coupling coefficients $\lambda$, $\lambda'$, $\mu_r$, $\mu'_r$, $\zeta'$, $\zeta_r$. The constraint $df/dt<0$ imposes that the symmetric part of the matrix of coupling coefficients (formally described in Eqs.\ref{coupling_array}, \ref{coeffs}) must be negative definite. This leads to sign constraints on the coupling coefficients present in Eq.\ref{constitutive_N} (see Supplementary Materials (SM) \cite{supp} section I). It is required in particular that $\xi>0$, $\xi_r>0$, $\xi_{||}>0$, $\xi'_r>0$, $\alpha>0$.

\section{Rotating cell doublets}
We now apply this generic theory to the case of a pair of cells in the low Reynolds number limit. Cell doublets have been shown to be able to spontaneously rotate \cite{brangwynne_symmetry_2000,lee_neighbor-enhanced_2021,lu_polarity-driven_2022} and we wish to understand broadly which fundamental physical mechanisms can lead to self-organised persistent rotation. We model cell dynamics using the simplified constitutive equations (Eq.\ref{constitutive_N}), and the following free energy terms:
\begin{equation}
\begin{aligned}
f_0\left( \bm p_i, \{N^{(i)}_k\} \right) &= \tilde{f}_0\left(\{N^{(i)}_k\}\right) + \kappa_i \bm p_i^2~,\\
f_{\text{int}}(\bm p_1,\bm p_2,\bm r_{12}, \{N_k^{(1)}\}, \{N_k^{(2)}\}) &= \kappa_r \bm r_{12}^2~,
\end{aligned}\label{free_energy_lagrange}
\end{equation}
where the Lagrange multipliers $\kappa_i$, $\kappa_r$ are introduced to enforce $||\bm p_1||=||\bm p_2||=1$ and $||\bm r_{12}||=2R$, with $R$ the cell radius, and $\tilde{f}_0$ determines the difference of chemical potential $\Delta\mu$.

In total, there are fourteen different parameters in this model, listed in Appendix \ref{dimensionless_appendix}. In the applications presented here, we set the active torque terms $\mu_r$ and $\mu'_r$ to zero. As a consequence, the torque balance equations impose $\bm \Omega_1=\bm \Omega_2$, such that the rotational friction term $\xi'_r(\bm \Omega_1-\bm \Omega_2)$ does not play a role, and the rotational friction coefficient $\xi'_r$ has no effect (Appendix \ref{3d_eq_appendix}, Eq.\ref{eq_xir_noeffect}){. We then take the limit of $\alpha\rightarrow \infty$, while imposing $||\bm p_i||=1$. The ten parameters that remain are the cell radius $R$, the active propulsion terms $\lambda$, $\lambda'$, the flow alignment terms $\eta$, $\eta'$, the polarity remodelling terms $\zeta_r$, $\zeta'$, and the friction terms $\xi$, $\xi_{||}$, $\xi_r$.}

\begin{figure}
\centering
\includegraphics[width=\linewidth]{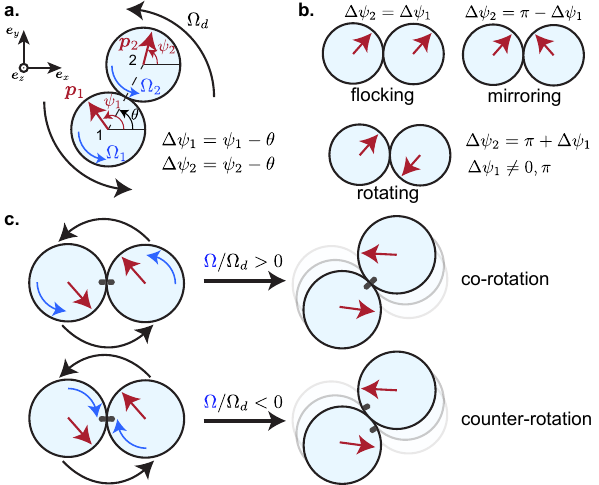}
\caption{\textbf{a.} Geometric quantities used to describe the motion of a cell doublet in the plane. \textbf{b.} Steady-states with different symmetries. \textbf{c.} Cases of doublet rotation: cells rotate with (top) or against (bottom) the doublet rotation. In the case of counter-rotation, constant polarity reorientation is necessary to maintain the angles $\Delta \psi_i$ constant. Here $\Omega=\Omega_1=\Omega_2$. $\Omega_i=\bm\Omega_i \cdot \bm e_z$ with $i=1,2$, $\Omega_d=\bm\Omega_d \cdot \bm e_z$.}\label{linear_stab}
\end{figure}

We focus on planar rotation for simplicity (Fig.\ref{linear_stab}a, Appendix \ref{section2D_appendix}).
The dynamics of the system in the 2D plane is reduced to a system of coupled differential equations for the angles $\Delta \psi_1(t)$, $\Delta \psi_2(t)$ and $\theta(t)$ defined in Fig. 2a:
\begin{align}
\frac{d\Delta \psi_1}{dt} & = F(\Delta \psi_1, \Delta \psi_2)~, \label{psi1} \\ 
\frac{d\Delta \psi_2}{dt} & = -F(\pi-\Delta \psi_2, \pi-\Delta \psi_1)~, \label{psi2} \\ 
\frac{d \theta}{dt} & = G(\Delta \psi_1, \Delta \psi_2)~, \label{theta}
\end{align}
with $F$ and $G$ given in Appendix \ref{2d_dynamics_FG_appendix}. The evolution of the angles $\
\Delta\psi_1$, $\Delta\psi_2$ is autonomous, and entirely determined by the properties of $F$. The overall doublet rotation rate rate $d\theta/dt$ is then obtained from $\Delta\psi_1$, $\Delta\psi_2$ and the function $G$. The velocity of the center of mass of the doublet can further be determined as a function of $\Delta \psi_1$, $\Delta\psi_2$ and $\theta$. We look for steady-state solutions of Eqs.\ref{psi1}, \ref{psi2} that have specific symmetries shown on Fig.\ref{linear_stab}b. These are the only steady-states (Appendix \ref{ss-sym}) in the cases studied in this paper (Appendix \ref{dimensionless_appendix}). We also define the ``in" state where polarities point towards each other ($\Delta\psi_1=0$, $\Delta\psi_2=\pi$) and the ``out" state where polarities point away from each other $\Delta\psi_1=\pi$, $\Delta\psi_2=0$, which are special cases at the intersection of ``rotating" and ``mirroring" states. We also note that the ``flocking'' and ``mirroring'' states coincide for $\Delta\psi_1=\Delta\psi_2=\pm \pi/2$.
By symmetry only the configuration $\Delta \psi_2=\pi+\Delta \psi_1$, $\Delta\psi_1\neq 0,\pi$ can lead to rotational motion of the doublet, $d\theta/dt \neq 0$. 
Permanent doublet rotation with angular velocity $\bm \Omega_d = \bm r_{12}\times (\bm v_2-\bm v_1)/(4 R^2)$ can be associated to various degrees of different intrinsic cell rotation $\bm \Omega_1, \bm \Omega_2$ (Fig. \ref{linear_stab}c). We study first how the combination of flow-alignment with active propulsion can lead to spontaneous rotation, and in a second part we show that rotation can also be observed with active propulsion combined with active polarity remodelling by adding one higher order term to the constitutive equations.
\subsection{Rotation from flow alignment}
In this part we restrict ourselves to passive polarity dynamics ($\zeta_r=\zeta'=0$) so that the only active forces that can drive rotation comes from either $\lambda$ or $\lambda'$. We then consider separately the case where motion is driven by cell-medium interactions ($\lambda$ and $\eta$ only) and cell-cell interactions ($\lambda'$ and $\eta'$ only). 

\begin{figure}
\centering
\includegraphics[width=\linewidth]{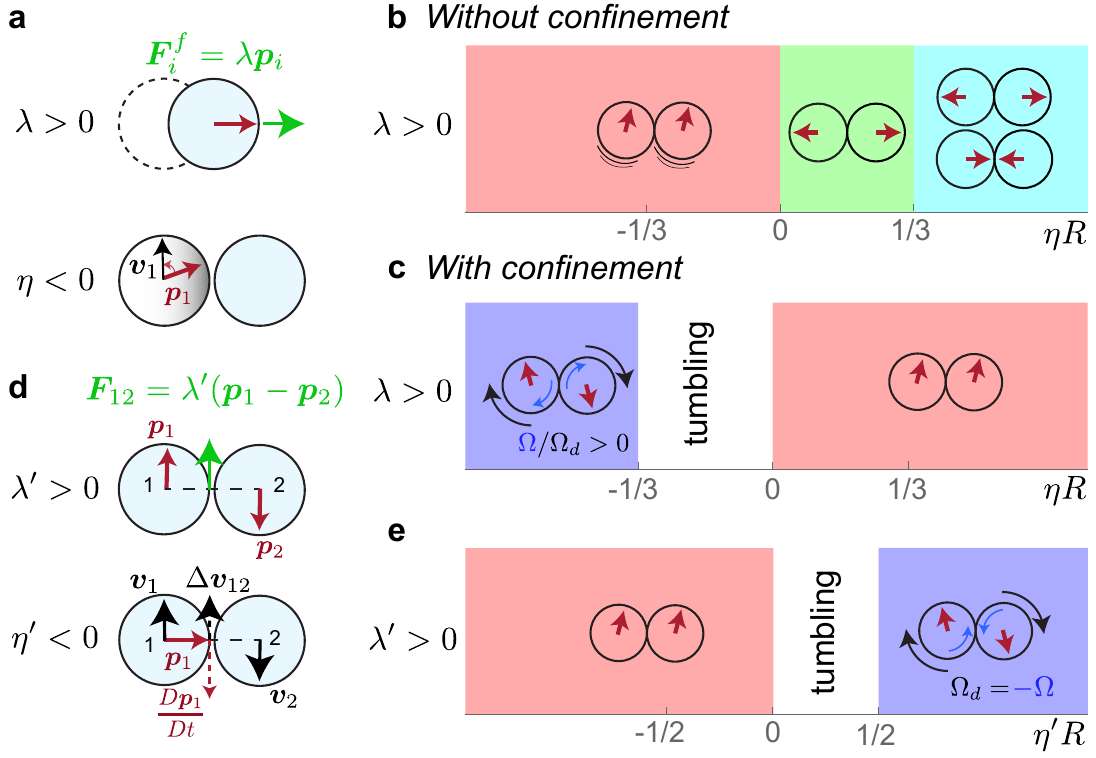}
\caption{\textbf{a}: Scheme explaining the effect of $\lambda$ (self-propulsion) and $\eta$ (flow alignment). \textbf{b, c}: Phase diagrams as a function of $\eta$, $\lambda$.  \textbf{d}: Scheme explaining the effect of $\lambda'$ (relative propulsion) and $\eta'$ (relative flow alignment). \textbf{e}: Phase diagram as a function of $\eta'$ and $\lambda'$. Red regions correspond to states of aligned polarities (motile in b, non motile in c,e. Purple regions correspond to rotating states. Green and cyan regions correspond respectively to ``in" and ``out" states where polarities point towards/away from each other. Marginally stable mirror states exist for all parameters in c, e. $\xi_r/(R^2\xi)=\xi_{||}/\xi=1$.}\label{phasediagetap}
\end{figure}

In the case of cell-medium interactions, we take $\lambda'=\eta'=0$, cells are self-propelled ($\lambda$) and align their polarity with their velocity vector ($\eta$). We make the system dimensionless (Appendix \ref{dimless-1}) and assume for simplicity that the dimensionless friction terms ($\xi_{||}/\xi$, $\xi_r/(\xi R^2)$) are equal to one, thus leaving only two dimensionless parameters, which are $\lambda/|\lambda|$ (the sign of the self-propulsion term) and $\eta R$. We find that the only non-static (marginally) stable steady-states in that case are the ``flocking" states ($\Delta\psi_1=\Delta\psi_2$, Fig.\ref{linear_stab}b) with any arbitrary angle $\Delta \psi_1$, for $\eta\lambda<0$ (Fig. \ref{phasediagetap}a,b). This is reminiscent of the flocking behaviour observed in large assemblies of self-propelled particles with flow alignment \cite{PhysRevE.74.061908}. For $\eta\lambda>0$, ``in" and ``out" states are the only possible stable states. Rotating states are unstable in this case.

Interestingly, if the doublet is confined (by pinning the center of mass using an external force, see Appendix \ref{2d_dynamics_FG_appendix}) the rotating solution is stabilised for $\lambda \eta<0$ and large enough $|\eta|$ (Fig.\ref{phasediagetap}c). Cells rotate in the {\it same} direction as the doublet, $\Omega/\Omega_d>0$. The complex interplay between self-propulsion, flow alignment and doublet rotation results in a strong effect of confinement on the doublet phase diagram. Indeed, confinement changes the cell velocities, which in turn impact  polarity dynamics through the flow alignment term in $\eta$. As a result, rotating states which are unstable without confinement can become stable (see SM section II for more details). In addition, stable static states with aligned polarities appear for $\lambda\eta>0$ (Fig. \ref{phasediagetap}c). Confinement can arise experimentally if cells are surrounded by elastic extracellular matrix \cite{lu_polarity-driven_2022} or on 2D substrates where only a patch of surface is adhesive \cite{brangwynne_symmetry_2000}. 

Stable rotating states also appear when cells actively move against each other ($\lambda'$) and have relative flow alignment ($\eta'$) but do not self-propel and do not orient their polarity according to their velocity relative to the fluid, $\lambda=\eta=0$ (Fig. \ref{phasediagetap}d, e). In this case, we consider the dimensionless parameters (Appendix \ref{dimless-2}) are $\lambda'/|\lambda'|$, $\eta'R$, $\xi_{||}/\xi$, $\xi_r/(\xi R^2)$. Active cell-cell interactions could arise in a biological tissue from the dynamics of adhesion molecules and actomyosin cortical flows at cell-cell interfaces. Here the surrounding medium only provides resistance to motion through translational and rotational frictions $\xi$ and $\xi_r$. Cells push or pull on each other and the doublet's center of mass does not move, hence confinement has no effect in that case. Here as well, the pink region of the phase diagrams in Fig. \ref{phasediagetap}e does not correspond to true ``flocking" states, but rather to static states with aligned polarities. The angular velocities of the doublet $\bm \Omega_d$ and cells $\bm \Omega_1$, $\bm \Omega_2$ then verify:
\begin{equation}
\bm \Omega_d = -\frac{\xi_r}{\xi R^2}\frac{\bm \Omega_1+\bm \Omega_2}{2}~.
\end{equation}
Since we have in addition $\bm \Omega_1=\bm \Omega_2=\bm \Omega$, $\bm \Omega_d = -(\xi_r/ (\xi R^2)) \bm \Omega$ and therefore cells are rotating in the \emph{opposite} direction to doublets (Fig.\ref{linear_stab}c).

In the white regions of the phase diagrams (Fig. \ref{phasediagetap}c,e), both the rotating and flocking states are unstable. Interestingly, mirror steady-states ($\Delta \psi_2=\pi-\Delta\psi_1$, Fig.\ref{linear_stab}b) exist for all values of $\Delta \psi_1$ but are unstable within a range of angles $\Delta \psi_1$. As a result, in the presence of noise, the system diffuses through the marginally stable mirror states until reaching the instability region, performing a ``tumble'', before diffusing again within marginally stable mirror states (SM section III.A). We note that in the presence of noise, transient diffusion across mirror states can occur for all parameters in the diagrams of Fig. \ref{phasediagetap}c,e, since marginally stable mirror states exist for all parameters.

\subsection{Rotation from active propulsion and polarity remodelling}\label{self-prop-section}
\begin{figure}
\centering
\includegraphics[width=\linewidth]{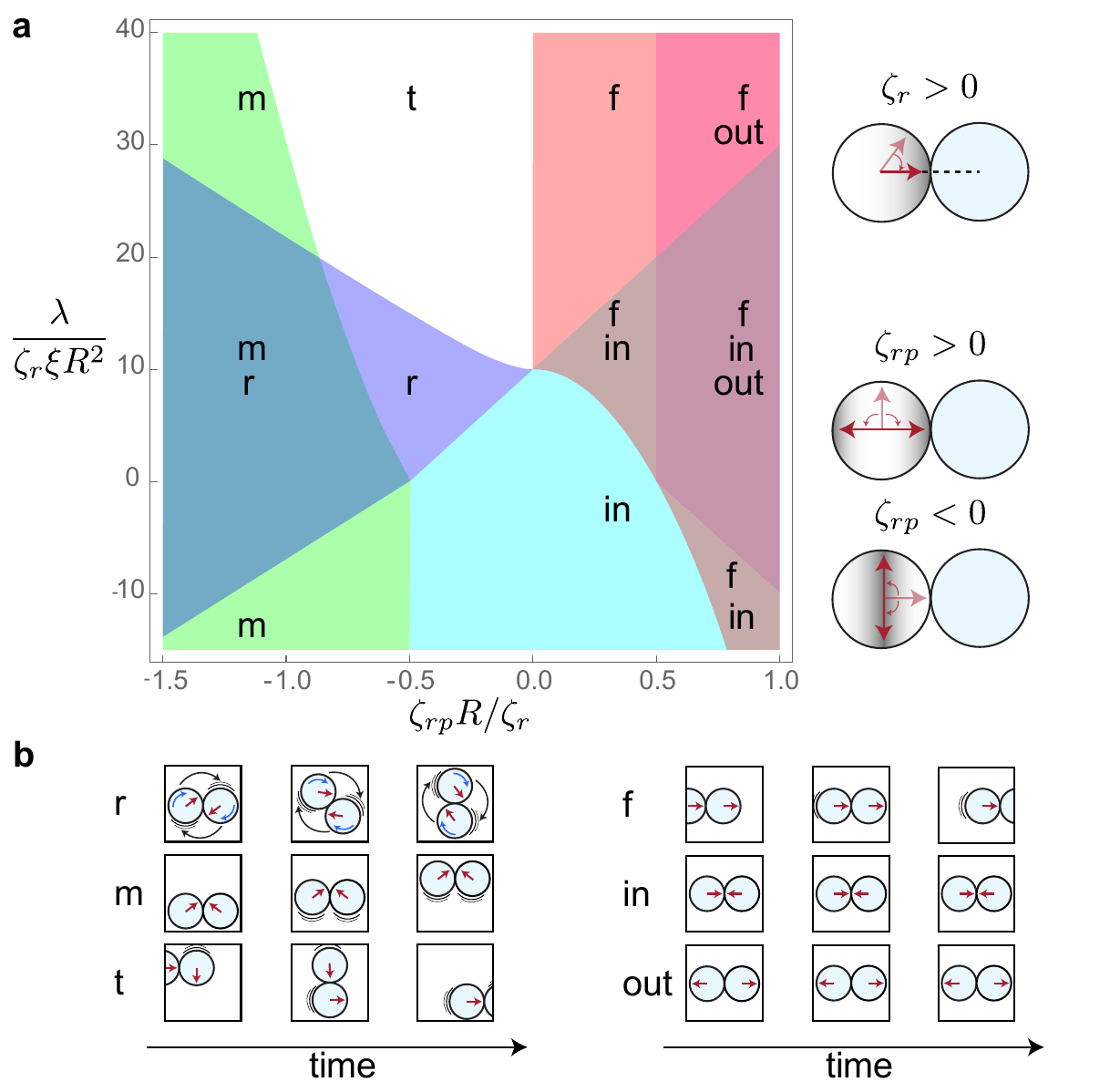}
\caption{\textbf{a:} Phase diagram of stable steady-states as a function of the dimensionless self-propulsion $\lambda/(\zeta_r\xi R^2)$ and the dimensionless polarity remodelling ratio $\zeta_{rp}R/\zeta_r$. Right, schematic for key couplings. Several stable steady-states coexist for the same parameters, when more than one label is written on a given region. \textbf{b:} Schematics for possible doublet dynamics: r (rotating), m (mirror), t (tumbling), f (flocking), in (polarities pointing towards each other), out (polarities pointing away from each other). By default, $\xi_r/(R^2\xi)=\xi_{||}/\xi=1$. For $\lambda=0$, the flocking and rotating states are non-motile. }\label{phasediag_zetarp}
\end{figure}

We now look for a case of doublet rotation that emerges from the interplay between self-propulsion and active polarity remodelling.  We consider a situation without flow alignment of polarity ($\eta=\eta'=0$), where cells self-propel along their polarity axis ($\lambda$) but not relative to each other, $\lambda'=0$, and where neighbouring cell polarities interact ($\zeta_r$, $\zeta'$). At this level of description, however, there are no rotating states, regardless of how $\lambda$, $\zeta_r$ and $\zeta'$ are combined. We therefore ask if higher order terms in the expansion of phenomenological coupling coefficients in $\bm p_i$ and $\bm r_{ij}$ would result in doublet rotation.
We therefore replace $\zeta_r \bm r_{ij}$ (in Eq.\ref{constitutive_N}) by:
\begin{equation}
\zeta_r \bm r_{ij} + \zeta_{rp}(\bm p_i \cdot \bm r_{ij}) \bm r_{ij}~.
\end{equation}
where we have added the next-order term in $\bm p_i$, $\bm r_{ij}$ which is not cancelled by the constraint of fixed polarity norm.
$\zeta_r$ (assumed positive) characterises the propensity of cells to orient their polarities towards each other, while $\zeta_{rp}$ (when negative) promotes an arrangement of polarities perpendicular to $\bm r_{ij}$ (Fig.\ref{phasediag_zetarp}a right).
For simplicity we set $\zeta'=0$ in the following, and the dimensionless parameters (see Appendix \ref{dimless-3}) are $\lambda/(\zeta_r \xi R^2)$, $\zeta_{rp}R/\zeta_r$,  $\xi_{||}/\xi$ and $\xi_r/(\xi R^2)$. $\lambda/(\zeta_r \xi R^2)$ is the ratio of time scales $1/(R\zeta_r)$ of polarity remodelling and $\xi R/|\lambda|$ of cell motion; and $\zeta_{rp}R/\zeta_r$ sets a preferred angle betwen $\bm p_i$ and $\bm r_{ij}$. The phase diagram of the system is rich and shows a multitude of steady-states (Fig.\ref{phasediag_zetarp}a, b) which sometimes coexist. In the regions of the diagram labelled ``f'' a ``flocking" state with $\Delta\psi_1=0$, corresponding to cells following each other, is stable. This state is reminiscent of contact following \cite{cf1,cf2,cf3,cf4}. A region exists for intermediate values of $\zeta_{rp}R/\zeta_r$ and large enough values of $\lambda/(\zeta_r \xi R^2)$ where this is the only stable state. Spontaneous rotation is possible provided that $\zeta_{rp}R/\zeta_r<0$ and for sufficiently low absolute values of $\lambda/(\zeta_r\xi R^2)$.
In that case the ratio between the cell and doublet angular velocities $\Omega$, $\Omega_d$ is:
\begin{equation}
\frac{\Omega}{\Omega_d}=\frac{2\xi_{||}}{\xi_r/R^2+2\xi_{||}}~.
\end{equation}
Here $\Omega$ and $\Omega_d$ have the same sign and therefore cells rotate in the \emph{same} direction than the doublet (Fig. \ref{linear_stab}c). When the cell-cell friction coefficient $\xi_{||}$ becomes large, $\Omega/\Omega_d\rightarrow 1$, the doublet rotates as a solid object, and the polarity simply co-rotates with the cell, which favors rotating steady-states (SM section IV). In the white region in the phase diagram of Fig.\ref{phasediag_zetarp}, the angles $\Delta \psi_1$, $\Delta \psi_2$ oscillate deterministically, the doublet moves in a mean direction with alternating phases of low velocity and high rotational motion, and larger velocity and low rotational motion (SM section III.B).
\section{Discussion}
We have developed a generic theory to describe the collective dynamics of aggregates of particles, described by their polarity $\bm p_i$, their velocity $\bm v_i$ and their angular velocity $\bm \Omega_i$. The framework of linear irreversible thermodynamics allows to derive systematically constitutive equations and to identify passive and active coupling coefficients. We distinguish between active self-propulsion terms, active interaction forces, and active interaction torques. In addition, explicitly introducing both the polarity $\mathbf{p}_i$ and angular velocity $\bm \Omega_i$ of particles allow to disentangle polarity dynamics from overall particle rotation. As a result, our physical description clarifies how mechanical interactions determine a particle's rotation, which in turn influence the polarity orientation.

Our study of cell doublets highlights the importance of individual cell rotation, which competes with active polarity remodelling (or flow polarity-alignment) to determine the possible configurations of cell doublets. 
We also find that counter or corotation of the cells participating to a rotating doublet (Fig. \ref{linear_stab}c) distinguishes two regimes whether rotation results from relative cell-cell propulsion (Fig. \ref{phasediagetap}e) or from self-propulsion (Fig. \ref{phasediagetap} c, Fig. \ref{phasediag_zetarp}).
We therefore argue that measuring cell rotation in experiments \cite{bajpai_interplay_2019} can allow to discriminate between models of cell and tissue dynamics. This would require to identify fiducial markers whose movement is well correlated with the overall cell rotation.


We find that a cell doublet can transition from a static ``in" state to a motile ``flocking" state when the active self-propulsion term $\lambda$ increases (Fig. \ref{phasediag_zetarp}a). This is in line with observations of cell doublets confined in a one dimensional channel which either polarise in the same direction at large velocity, or polarise in opposite directions when they are less motile \cite{zhang_morphodynamic_2021}. Our model also predicts the existence of parameter regimes where cells alternate between phases of quick reorganisation of their polarities (tumbles) and phases of slow noise driven diffusive dynamics (Fig. \ref{phasediagetap}c, e) or alternate between tumbles and ``runs" where the doublet translates as a whole (Fig. \ref{phasediag_zetarp}a). This  behaviour is reminiscent of experimental observations of clusters of 50-100 cells collectively show phases of rotation, translation or random organisation  \cite{malet-engra_collective_2015}. We expect that our framework will find applications to describe the dynamics of larger aggregates with $N>2$ cells.

Our theory, by describing cells as active polar spinning particles, obviously relies on strong simplifications. We limited ourselves to the description of cells as rigid bodies, but it would be interesting to extend this framework to non-rigid body motions. Coarse-grained modes of cell deformation could be treated as additional degrees of freedom. We have used linear irreversible thermodynamics to identify coarse-grained fluxes and forces and obtain linear relations between them. Since these relations apply strictly speaking close to equilibrium, comparison to experiments will determine if additional non-linear couplings or coarse-grained variables must be introduced. Coupling coefficients breaking Onsager reciprocity relations could also be investigated \cite{bowick2022symmetry}.
 We have included a single chemical reaction fuelling active processes. It would be interesting to consider additional chemical reactions, and to distinguish molecular species located in bulk and at interfaces between cells. To describe fluctuations, noise terms could be added to the expansion of fluxes into forces, Eq. \ref{eqlinearresponse}, as for instance described as Ref. \cite{julicher_hydrodynamic_2018}. Our theory could be expanded to introduce chiral effects such as an active torque $\bm \Gamma_i^f$ proportional to the cell polarity $\bm p_i$ \cite{liebchen2022chiral}. 

\begin{acknowledgments}
We thank T. Guyomar, L. Lu and D. Riveline for interesting discussions on doublet rotation, A. Torres-Sanchez for pointing out Ref. \cite{admal_unified_2010}, M. Kothari for feedback and K. Kruse for critical reading of the manuscript.
\end{acknowledgments}

\appendix
\section{Mechanics of interacting macroscopic objects}\label{micro-macro-appendix}
We consider here the case of an ensemble of macroscopic physical objects, consisting of interacting point particles. The word ``particle'', in this section, refers to the microscopic constituents of macroscopic objects. We introduce the interparticle potential energy $U$ which is taken to depend only on the positions of the particles $\bm r_k$:
\begin{equation}
U(\{\bm r_k\}_{\text{all}})~,
\end{equation}
where $\{\bm r_k\}_{\text{all}}$ denotes the set of the positions of all particles.  $U$ is assumed to be invariant by a uniform translation and rotation, which leads to:
\begin{equation}
\sum_{k\in \text{all}} \frac{\partial U}{\partial \bm r_k} = 0~,~\sum_{k \in \text{all}} \frac{\partial U}{\partial \bm r_k} \times \bm r_k  = 0~.\label{transrotinv}
\end{equation}
The total force $\bm f_k$ exerted on particle $k$ is:
\begin{equation}
\bm f_k = - \frac{\partial U}{\partial \bm r_k}~.
\end{equation}
Introducing $m_k$ the mass of particle $k$, the second law of Newton is:
\begin{equation}
m_k \frac{d\bm v_k}{dt} = \bm f_k ~.
\end{equation}
For a general potential $U$ satisfying invariance by uniform translations and rotations, not necessarily consisting of pairwise interactions, it can be shown \cite{admal_unified_2010} that the force $\bm{f}_k$ acting on a particle can be decomposed into a sum of central forces, involving effective interaction forces $\bm f_{kl}$ that verify the following properties:
\begin{equation}
\forall k~\bm f_k  = \sum_{l \in \text{all}} \bm f_{kl}~,~
\forall k,l~\bm f_{kl} = - \bm f_{lk}~,~
\forall k,l~\bm f_{kl} \propto \bm r_k - \bm r_l~,\label{micro_properties}
\end{equation}
and this decomposition is not unique. 

We now consider an ensemble of macroscopic objects (``cells'') within a fluid, and consider each of them as a set of point particles. We therefore introduce disjoint sets of particles $\mathbb{C}_1$, $\mathbb{C}_2$, ... , $\mathbb{C}_{n_c}$, and $\mathbb{F}$ corresponding to the $n_c$ cells and the surrounding fluid. The center of mass $\bm R_i$ of cell $i$ is defined by:
\begin{equation}
\left(\sum_{k \in \mathbb{C}_i} m_k \right)\bm R_i =m_i \bm R_i = \sum_{k \in \mathbb{C}_i} m_k \bm r_k~,
\end{equation}
where we have introduced $m_i=\sum_{k \in \mathbb{C}_i} m_k $ the mass of cell $i$.
The center of mass velocity of cell $i$ is given by
\begin{align}
\bm{v}_i=\frac{d\bm R_i}{dt}~.
\end{align}
A macroscopic version of Newton's second law can be derived for the center of mass:
\begin{equation}
m_i \frac{d\bm v_i}{dt} = \sum_{k \in \mathbb{C}_i} \sum_{l \neq k} \bm f_{kl} = \bm F_i~.
\end{equation}
By using the property $\bm f_{kl}=-\bm f_{lk}$, the interactions between particles of $\mathbb{C}_i$ vanish and we can split $\bm F_i$ into different contributions:
\begin{equation}
\begin{split}
\bm F_i =  \bm F_i^f + \sum_{j\neq i}\bm F_{ij} \text{ with } \bm F_i^f = \sum_{k \in \mathbb{C}_i} \sum_{l \in \mathbb{F}}\bm f_{kl}\\
  \text{ and } \bm F_{ij}=\sum_{k \in \mathbb{C}_i} \sum_{l \in \mathbb{C}_j}\bm f_{kl}~.
\end{split}
\end{equation}
Furthermore, the property $\bm f_{kl}=-\bm f_{lk}$ extends to the macroscopic interaction forces and we find the macroscopic third law of Newton:
\begin{equation}
\bm F_{ij} = - \bm F_{ji}~. \label{3rdlawforce}
\end{equation}
The spin angular momentum of cell $i$ is defined as:
\begin{equation}
\bm L_i =\sum_{k \in \mathbb{C}_i}  (\bm r_k - \bm R_i)\times (m_k \bm v_k)~,
\end{equation}
and its time derivative obeys:
\begin{equation}
\frac{d \bm L_i}{dt} = \sum_{k \in \mathbb{C}_i}  \sum_{l \neq k} (\bm r_k - \bm R_i)\times \bm f_{kl} = \bm \Gamma_i~,
\end{equation}
where $\bm \Gamma_i$ is the torque acting on cell $i$, relative to its center of mass. Again we can split the torque in different contributions:
\begin{multline}
\label{SM:eq_Gamma_i_decomposition}
\bm \Gamma_i = \sum_{k \in \mathbb{C}_i} \sum_{l \in \mathbb{C}_i}  \bm r_{lk}\times \bm f_{kl} \\
+\sum_{j\neq i}\left( \sum_{k \in \mathbb{C}_i} \sum_{l \in \mathbb{C}_j} (\bm r_k - \bm R_i) \times \bm f_{kl}\right) \\
+ \sum_{k \in \mathbb{C}_i} \sum_{l \in \mathbb{F}} (\bm r_k - \bm R_i) \times \bm f_{kl}~,
\end{multline}
where $\bm r_{kl}=\bm r_l - \bm r_k$ is the distance between two particles. Because $\bm f_{kl} \propto\bm r_{kl}$, the first term in the right hand side of Eq. \ref{SM:eq_Gamma_i_decomposition} vanishes (corresponding to no internal torque generation), and we obtain:
\begin{multline}
\bm \Gamma_i  = \bm \Gamma_i^f + \sum_{j \neq i}\bm \Gamma_{ij}~,~ \\
 \text{with } \bm \Gamma_i^f  = \sum_{k \in \mathbb{C}_i} \sum_{l \in \mathbb{F}} (\bm r_k - \bm R_i) \times \bm f_{kl}  ~,~ \\
 \text{and }  \bm \Gamma_{ij}  = \sum_{k \in \mathbb{C}_i} \sum_{l \in \mathbb{C}_j} (\bm r_k - \bm R_i) \times \bm f_{kl}~.
\end{multline}
An extension of Newton's third law to macroscopic torques is obtained by computing:
\begin{equation}
\bm \Gamma_{ij}+\bm \Gamma_{ji} =  \sum_{k \in \mathbb{C}_i} \sum_{l \in \mathbb{C}_j}\Bigl[(\bm r_k - \bm R_i)\times \bm f_{kl} + (\bm r_l-\bm R_j)\times \bm f_{lk}\Bigr]~.
\end{equation}
Using the properties $\bm f_{kl}=-\bm f_{lk}$ and $\bm f_{kl}\propto \bm r_{kl}$, we find that for any pair $(i,j)$ of cells:
\begin{equation}
\bm \Gamma_{ij}+\bm \Gamma_{ji} = \bm r_{ij} \times \bm F_{ij}\text{ with }\bm r_{ij}=\bm R_j - \bm R_i~. \label{3rdlawtorque}
\end{equation}
Because of this relation, we see that in general we can split the torque between two cells into a symmetric and an anti-symmetric part:
\begin{equation}
\bm \Gamma_{ij} = \frac{1}{2}\bm r_{ij}\times \bm F_{ij} + \bm \Gamma_{ij}^a\text{ with }\bm \Gamma_{ij}^a = -\bm \Gamma_{ji}^a~.\label{gamma_a}
\end{equation}
\section{Equilibrium forces and torques}\label{eq_fo_to}
Here we obtain equilibrium forces and torques by considering two interacting particles $1$, $2$ and a surrounding fluid, described by the free energy of Eq.\ref{free_energy}, for $N=2$:
\begin{multline}
\label{SM_eq_free_energy_definition}
f= \sum_{i=1}^{2}\left( \frac{1}{2}m\bm v_i^2+ \frac{1}{2} \bm L_{i}  \cdot \bm \Omega_{i}   \right)\\
+\sum_{i=1}^{2} f_0\left( \bm p_i\right) + f_{\text{int}}(\bm p_1,\bm p_2,\bm r_{12})~.
\end{multline}
We did not write the dependencies in $\{N^{(1)}_k\}$ and $\{N^{(2)}_k\}$ for simplicity, since they do not modify the computation of equilibrium torques and forces. The free energy $f_{\text{int}}$ does not depend on the order of labelling of particles $1$, $2$, such that $f_{\text{int}}(\mathbf{p}_2,\mathbf{p}_1,\mathbf{r}_{21})=f_{\text{int}}(\mathbf{p}_1,\mathbf{p}_2,\mathbf{r}_{12}$).
At equilibrium, we ignore the kinetic part of the free energy. We compute the change of free energy with respect to infinitesimal translations ($\delta \bm r_i$), rotations ($\delta \bm \theta_i$) and polarity changes ($D\bm p_i$) for each particle. Since rotating a particle also rotates its polarity, the total polarity change $\delta \bm p_i$ is given by:
\begin{equation}
\delta \bm p_i = \delta \bm \theta_i \times \bm p_i + D\bm p_i~,\label{dp}
\end{equation} 
where $D\bm p_i$ represents the polarity change that can be imposed independently from the rotation of the particles. 

The energy change $\delta f$ for a small deformation is equal to the work of the external forces on the system.  First, we apply this principle to the surrounding fluid. The free energy of the fluid does not depend on the particles positions and orientations. Therefore the infinitesimal virtual work done by external force $-\bm F^{f\text{eq}}_i$ and external torques $-\bm \Gamma^{f\text{eq}}_i$ exerted by the particles vanish:
\begin{equation}
0 = -\sum_{i=1}^2 \bm F^{f\text{eq}}_i\cdot \delta \bm r_i - \sum_{i=1}^2 \bm \Gamma^{f\text{eq}}_i \cdot \delta \bm \theta_i~.\label{fluid_work}
\end{equation}
Since this must be true for any $\delta \bm r_i$, $\delta \bm \theta_i$, we deduce that:
\begin{equation}
\bm F^{f\text{eq}}_i=0\text{ and }\bm \Gamma^{f\text{eq}}_i=0~.
\end{equation} 
Now we apply this principle to a system of two particles, excluding the surrounding fluid. The virtual work arising from arbitrary external forces $ \bm F_i$ and torques $\bm \Gamma_i$ on the particles can be rewritten, using the force balance equations $\bm F_1+\bm F_{12}=0$, $\bm F_2+\bm F_{21}=0$ and torque balance equations $\bm \Gamma_1+\bm \Gamma_{12}=0$, $\bm \Gamma_2 +\bm \Gamma_{21}=0$:
\begin{multline}
\sum_{i=1}^2\left( \bm F_i\cdot \delta \bm r_i + \bm \Gamma_i\cdot \delta \bm \theta_i \right)=\bm F_{12} \cdot (\delta \mathbf{r}_2-\delta\bm r_1) \\
- \bm \Gamma_{12} \cdot \delta \bm \theta_1 -  \bm \Gamma_{21} \cdot\delta \bm \theta_2~.
\end{multline}
At equilibrium the work exerted on the system is equal to its change of free energy, leading here to the relation:
\begin{multline}
\bm F_{12}^{ eq} (\delta \mathbf{r}_2-\delta\bm r_1) - \bm \Gamma_{12}^{ eq}  \delta \bm \theta_1 -  \bm \Gamma_{21} \delta \bm \theta_2\\
= \delta \left( \sum_{i=1}^{2} f_0\left( \bm p_i \right) + f_{\text{int}}(\bm p_1,\bm p_2,\bm r_{12}) \right)~.\label{work_2_cells}
\end{multline}
To compute the right hand side of the equation, we make use of the terms $\bm h_i^{(0)}$ and $\bm h_k^{\langle ij \rangle}$ defined in Eq.\ref{hdef}, and we define $\tilde{\bm F}_{ij}$ as:
\begin{equation}
\tilde{\bm F}_{ij}  = \frac{\partial f_{\text{int}}}{\partial \bm r_{ij}}(\bm p_i,\bm p_j,\bm r_{ij})~,
\end{equation}
where $\langle ij\rangle$ denotes pairs $1,2 $ or $2,1$.
Because $f_0$ and $f_{\text{int}}$ are invariant by rotation, we have:
\begin{equation}
\bm h^{(0)}_i \times \bm p_i = 0~,~ 
\bm h^{\langle ij \rangle}_{i}  \times \bm p_i +\bm h^{\langle ij \rangle}_{j} \times \bm p_j  = -\bm r_{ij} \times  \tilde{\bm F}_{ij}~.
\end{equation}
We also note that $\bm h^{\langle ij \rangle}_{k}=0$ for $k \notin \{i,j\}$, and by symmetry we have $\bm h^{\langle ij \rangle}_{k}=\bm h^{\langle ji \rangle}_{k}$. Eq.\ref{work_2_cells} then becomes:
\begin{multline}
\bm F_{12}^{ eq} (\delta \mathbf{r}_2-\delta\bm r_1) - \bm \Gamma_{12}^{eq}  \delta \bm \theta_1 -  \bm \Gamma_{21} \delta \bm \theta_2=\\
 \tilde{\bm F}_{12}\cdot (\delta \bm r_2 -\delta \bm r_1)-\sum_{i=1}^2\bm h_i \cdot \delta \bm p_i~.
\end{multline}
After using the force and torque balance equations, as well as introducing $D\bm p_i$, we obtain:
\begin{multline}
(\bm F_{12}^{eq}-\tilde{\bm F}_{12})\cdot(\delta \bm r_2 -\delta \bm r_1)+ \delta \bm \theta_1 \cdot ( -\bm h_1^{\langle 12\rangle}\times \bm p_1 -  \bm \Gamma_{12}^{eq} ) \\
+ \delta \bm \theta_2 \cdot ( -\bm h_2^{\langle 12\rangle}\times \bm p_2 -  \bm \Gamma_{21}^{eq} )+ \sum_{i=1}^2 \bm h_i D\bm p_i=0 ~.
\end{multline}
Since this must be true for every $\delta \bm r_i$, $\delta \bm \theta_i$, $D\bm p_i$, we deduce that for two interacting particles 1,2 (and by extension for two interacting particles $i$, $j$) the equilibrium forces and torques are:
\begin{equation}
\begin{aligned}
\bm F_{ij}^{eq} & = \tilde{\bm F}_{ij} = \frac{\partial f_{\text{int}}}{\partial \bm r_{ij}}~, \\
\bm \Gamma_{ij}^{eq} & = -\bm h^{\langle ij\rangle}_i \times \bm p_i = \frac{\partial f_{\text{int}}}{\partial \bm p_i}(\bm p_i,\bm p_j,\bm r_{ij})\times \bm p_i~,
\end{aligned}\label{eq_tor_for}
\end{equation}
In addition, the term in factor of $D\bm p_i$ shows that at equilibrium the condition $\bm h_i=0$ is satisfied.
\section{Time derivative of the kinetic energy}\label{kinetic-section}
Here we discuss the kinetic energy and its time derivative using classical notions of rigid body mechanics \cite{Landau1976Mechanics}. The kinetic energy is given by the first term in the right hand side of Eq. \ref{free_energy}.
For the translational part of the kinetic energy, the force balance equation \ref{force_bal} directly leads to:
\begin{equation}
\frac{d}{dt}\left(\frac{1}{2}m_i\bm v_i^2\right) = \bm F_i\cdot \bm v_i~. \label{kin_trans}
\end{equation}

We now discuss the time derivative of the rotational part of the kinetic energy.
The kinetic energy in Eq.\ref{free_energy} is obtained assuming that particles represent rigid bodies such that the velocity field $\bm v$ at at point $\bm r$ within particle $i$ can be written as:
\begin{equation}
\label{eq:velocity_within_particle}
\bm v = \bm v_i + \bm \Omega_i \times (\bm r - \bm r_i) \text{ with } m_i \bm r_i = \int_{V_i}dV \rho \bm r~.
\end{equation}
with $\rho$ the mass density, and the mass of the particle $i$ is given by:
\begin{equation}
\begin{aligned}
m_i & =\int_{V_i}dV\rho.
\end{aligned}
\end{equation}

The angular momentum of object $i$ around its center of mass and its time derivative obey the following equations \cite{Landau1976Mechanics}:
\begin{equation}
\begin{aligned}
\bm L = &\int dV (\bm r-\bm r_i)\times \rho \bm v~,\\
L_{i\alpha} =& I_{i \alpha \beta}\Omega_{i \beta} ~,
\end{aligned}
\end{equation}
where we use greek indices for cartesian coordinates of space and Einstein convention for summation of these indices, and the expression for the tensor of inertia $\bm I$ can be found using Eq. \ref{eq:velocity_within_particle}:
\begin{equation}
\begin{aligned}
I_{i \alpha \beta} & = \int_{V_i}dV\rho (\Delta r_{i\gamma}\Delta r_{i\gamma} \delta_{\alpha \beta} - \Delta r_{i \alpha} \Delta r_{i \beta} )~,
\end{aligned}
\end{equation}
with $\Delta \bm r_i = \bm r-\bm r_i$ .
If we compute the time derivative of the tensor of inertia, we find:
\begin{multline}
\frac{d I_{i\alpha \beta}}{dt} = -\int_{V_i}dV\rho\left[(\bm \Omega_i \times \bm \Delta r_i)_\alpha \Delta r_{i\beta}\right.\\
\left. + (\bm \Omega_i \times \bm \Delta r_i)_\beta \Delta r_{i\alpha}\right]~,
\end{multline}
where we have used that the mass density $\rho$ is conserved.
This simplifies to:
\begin{equation}
\frac{d I_{i\alpha \beta}}{dt} =\Omega_{i\alpha '} ( \epsilon_{\alpha' \beta ' \alpha}I_{\beta' \beta} +  \epsilon_{\alpha' \beta ' \beta}I_{\beta' \alpha}) \label{dIdt}~.
\end{equation}
We note here that this time derivative obeys the following property:
\begin{equation}
\Omega_{i\alpha} \frac{d I_{i\alpha \beta}}{dt} \Omega_{i\beta} = 0 \label{property_I}~.
\end{equation}
The time derivative of the rotational part of the kinetic energy can be expressed in two different ways:
\begin{equation}
\begin{aligned}
\frac{d}{dt}\left( \frac{1}{2}\Omega_{i\alpha} I_{i\alpha \beta} \Omega_{i\beta} \right) & = \frac{d \Omega_{i\alpha}}{dt}I_{i\alpha \beta}\Omega_{i\beta} + \frac{1}{2}\Omega_{i \alpha} \frac{dI_{i\alpha \beta}}{dt}\Omega_{i\beta} \\
& = \frac{d \Omega_{i\alpha}}{dt}L_{i\alpha} + \frac{1}{2}\Omega_{i \alpha} \frac{dI_{i\alpha \beta}}{dt}\Omega_{i\beta} \\
\frac{d}{dt}\left( \frac{1}{2}\Omega_{i\alpha} I_{i\alpha \beta} \Omega_{i\beta} \right) & = \frac{1}{2}\frac{d\Omega_{i\alpha}}{dt} L_{i\alpha} + \frac{1}{2} \Omega_{i\alpha} \frac{d L_{i\alpha}}{dt}
\end{aligned}
\end{equation}
Using Eq.\ref{property_I}, we deduce that $\frac{d \bm \Omega_i}{dt}\cdot \bm L_{i}$ = $\bm \Omega_i \cdot \frac{d \bm L_i}{dt}$ and therefore we have:
\begin{equation}
\frac{d}{dt}\left( \frac{1}{2}\Omega_{i\alpha} I_{i\alpha \beta} \Omega_{i\beta} \right) = \bm \Omega_{i} \cdot \frac{d \bm L_i}{dt} = \bm \Omega_i \cdot \bm \Gamma_i~,\label{kin_torq}
\end{equation}
where in the last equality  we have used the torque balance equation \ref{torque_bal}.
\onecolumngrid
\section{Detailed derivation of constitutive equations}\label{consteq_appendix}
It is useful to supplement the array of fluxes and forces (Table \ref{signatures}) by adding the time signatures and symmetry properties of each flux/force pair (N/A indicates non applicable):
\begin{equation}
\begin{array}{!{\vrule width 1pt} c | c|  c | c| c !{\vrule width 1pt}}
\Xhline{1pt}
 \text{Force} & \text{Flux} & \text{Time signature (flux)} & \text{Parity} & i,j \text{ switch}  \\
 \Xhline{1pt}
 \Delta \mu_i & r_i & -1 & \text{scalar} & N/A \\
  \hline
 -\bm h _i & \frac{D \bm p_i}{Dt} & -1 & \text{vector} & N/A \\
 \Xhline{1pt}
  \bm \Omega_i & \bm \Gamma^f_i & 1 & \text{pseudovector} & N/A \\
 \hline
 \bm v_i & \bm F^f_i & 1 & \text{vector} & N/A \\
 \Xhline{1pt}
 \Delta \bm v_{ij} & \bm F_{ij}^d & 1 & \text{vector} & -1 \\
 \hline
 \bm \Omega_i - \bm \Omega_j & \bm \Gamma_{ij}^d & 1 & \text{pseudovector} & -1 \\
 \Xhline{1pt}
\end{array} \label{signatures_detailed}
\end{equation}
In addition to Eq.\ref{onsager_symmetry}, the coupling coefficients must also be consistent with the fact that fluxes and forces are scalar, vector, or pseudo-vector quantities. Considering a generic scalar flux $J$ (or vector $\bm J$, or pseudo-vector $\tilde{\bm J}$), and a generic scalar force $F$ (or vector $\bm F$, or pseudo-vector $\tilde{\bm F}$), possible couplings are:
\begin{equation}
\begin{aligned}
J&=OF~,~J= \bm O \cdot \bm F~,~J=\tilde{\bm O}\cdot \tilde{\bm F}~, \\
\bm J & = \bm O F~,~\bm J = \underline{\bm O} \bm F~,~\bm J =\tilde{\underline{\bm{O}}} \tilde{\bm F}~, \\
\tilde{\bm J} &= \tilde{\bm O} F~,~\tilde{\bm J} = \tilde{\underline{\bm{O}}} \bm F ~,~\tilde{\bm J} = \underline{\bm O} \tilde{\bm F}~.
\end{aligned}\label{vector-pseudovector-symmetry}
\end{equation}
We have introduced scalar ($O$), vector/pseudo-vector ($\bm O$, $\tilde{\bm O}$), and tensor/pseudotensor ($\underline{\bm O},\tilde{\underline{\bm O}}$) coupling coefficients. Under the assumptions introduced in section \ref{consteq_section_main}, we obtain the following coupling table:
\begin{equation}
\begin{aligned}
r_i  = &~ O_{11}^i \Delta \mu_i + \sum_{j\in \mathcal{N}_i}O_{11}^{ij}\Delta\mu_j-\bm{O}^i_{12}\cdot \bm h_i-\sum_{j\in \mathcal{N}_i}\bm O_{12}^{ij}\cdot \bm h_j+\tilde{\bm O}^i_{13}\cdot \bm \Omega_i+\bm O^i_{14}\cdot \bm v_i+\sum_{j\in \mathcal{N}_i}\bm O_{15}^{ij}\cdot \Delta\bm v_{ij}\\
& +\sum_{j\in \mathcal{N}_i}\tilde{\bm O}_{16}^{ij}\cdot(\bm \Omega_i - \bm \Omega_j),\\
\frac{D\bm p_i}{Dt} = & ~ \bm O_{12}^i\Delta \mu_i+\sum_{j\in \mathcal{N}_i}\bm O_{12}^{ji}\Delta \mu_j-\underline{\bm O}_{22}^i\bm h_i-\sum_{j\in \mathcal{N}_i}\underline{\bm O}_{22}^{ij}\bm h_j+\underline{\tilde{\bm O}}_{23}^i\bm \Omega_i+\underline{\bm O}_{24}^i\bm v_i+\sum_{j\in \mathcal{N}_i}\underline{\bm O}_{25}^{ij}\Delta \bm v_{ij}+\sum_{j\in \mathcal{N}_i}\underline{\tilde{\bm O}}_{26}^{ij}(\bm \Omega_i-\bm \Omega_j), \\
\bm \Gamma_i^f = & ~ -\tilde{\bm O}_{13}^i\Delta \mu_i+{}^t\underline{\tilde{\bm O}}_{23}^i\bm h_i+\underline{\bm O}_{33}^i\bm \Omega_i+\tilde{\underline{\bm O}}_{34}^i\bm v_i,\\
\bm F_i^f = & ~ -\bm O_{14}^i\Delta \mu_i+{}^t\underline{\bm O}_{24}^i \bm h_i+{}^t\underline{\tilde{\bm O}}_{34}^i\bm \Omega_i+\underline{\bm O}_{44}^i \bm v_i,\\
\bm F_{ij}^d = & ~ - \bm O_{15}^{ij}\Delta\mu_i+\bm O_{15}^{ji}\Delta\mu_j+{}^t\underline{\bm O}_{25}^{ij}\bm h_i-{}^t\underline{\bm O}_{25}^{ji}\bm h_j+\underline{\bm O}_{55}^{ij}\Delta \bm v_{ij}+\tilde{\underline{\bm O}}_{56}^{ij} (\bm \Omega_i-\bm \Omega_j),\\
\bm \Gamma_{ij}^d = & -\tilde{\bm O}_{16}^{ij}\Delta \mu_i+\tilde{\bm O}_{16}^{ji}\Delta\mu_j+{}^t\tilde{\underline{\bm O}}_{26}^{ij}\bm h_i-{}^t\tilde{\underline{\bm O}}_{26}^{ji}\bm h_j+{}^t\tilde{\underline{\bm O}}_{56}^{ij}\Delta \bm v_{ij}+\underline{\bm O}_{66}^{ij}(\bm \Omega_i-\bm \Omega_j)~.
\end{aligned} \label{coupling_array}
\end{equation}
Here, ${}^t\underline{\bm A}$ is the transposed tensor of tensor $\underline{\bm A}$. Eq.\ref{coupling_array}, with the additional requirements that $O_{11}^{ij}=O_{11}^{ji}$, $\underline{\bm O}_{22}^{ij}={}^t \underline{\bm O}_{22}^{ji}$, $\underline{\bm O}_{55}^{ij}=\underline{\bm O}_{55}^{ji}$, $\underline{\tilde{\bm O}}_{56}^{ij}=\underline{\tilde{\bm O}}_{56}^{ji}$, $\underline{\bm O}_{66}^{ij}=\underline{\bm O}_{66}^{ji}$, and $\underline{\bm O}_{22}^i$, $\underline{\bm O}_{33}^i$, $\underline{\bm O}_{44}^i$, $\underline{\bm O}_{55}^i$, $\underline{\bm O}_{66}^i$ are symmetric, verifies the Onsager reciprocity relations (Eq.\ref{onsager_symmetry}), the constraints of Eq.\ref{vector-pseudovector-symmetry}, and respects the behaviour of $\bm F_{ij}^d$, $\bm\Gamma_{ij}^d$, $\Delta\bm v_{ij}$ and $\bm \Omega_i-\bm \Omega_j$ when the indices $i$ and $j$ are switched. 
We now need to find the expressions of the coupling coefficients, at the lowest order in $\bm r_{ij}$, $\bm p_i$, and number of molecules $N_k^i$. Doing so consistently with the assumptions introduced in section \ref{consteq_section_main}, we obtain the following expressions:
\begin{equation}
\begin{aligned}
\bm O_{12}^i & = \zeta_0 \bm p_i, \bm O_{12}^{ij}=(\zeta-\zeta_0)\bm p_j+\zeta'\bm p_i+\zeta_r\bm r_{ji},~\tilde{\bm O}^i_{13} = 0,~\bm O^i_{14}=-\lambda \bm p_i ,\\
\bm O^{ij}_{15}& =-\frac{\lambda'}{2}(\bm p_i-\bm p_j)+\frac{\lambda''}{2}(\bm p_i+\bm p_j)-\frac{1}{2}\lambda'_r \bm r_{ij},
\\ \tilde{\bm O}^{ij}_{16}& =-\frac{\mu_r}{2}\bm r_{ij}\times (\bm p_i+\bm p_j) +\frac{\mu''_r}{2}\bm r_{ij}\times (\bm p_i - \bm p_j)- \frac{\mu'_r}{2}\bm p_i\times \bm p_j,\\
O_{11}^i &= -\Lambda,O_{11}^{ij} = -\Lambda',~\underline{\bm O}_{22}^{i}=-\alpha\underline{\bm I}_d,~\underline{\bm O}_{22}^{ij}=-\alpha'\underline{\bm I}_d,~\underline{\bm O}_{33}^i=-\xi_r\underline{\bm I}_d,~\underline{\bm O}_{44}^i=-\xi\underline{\bm I}_d,~\underline{\bm O}_{55}^{ij}=-\xi_{||}\underline{\bm I}_d,~\underline{\bm O}_{66}^{ij}=-\xi'_r\underline{\bm I}_d,\\
\underline{\tilde{\bm O}}_{23}^i & =\chi \bm p_i \times,~\underline{\bm O}_{24}^i=-\eta\underline{\bm I}_d,~\underline{\bm O}_{25}^{ij}=\eta'\underline{\bm I}_d,~\underline{\tilde{\bm O}}_{26}^{ij} = \Bigl[ \gamma \bm p_i+\gamma' \bm p_j +\gamma'' \bm r_{ij}\Bigr] \times,\\
\underline{\tilde{\bm O}}^i_{34} & = \beta \bm p_i \times ,~\underline{\tilde{\bm O}}_{56}^{ij}=\Bigl[\beta'(\bm p_i+\bm p_j)\Bigr]\times~.\label{coeffs}
\end{aligned}
\end{equation}
where $\underline{\bm I}_d$ is the identity tensor, and pseudo-tensors are constructed as $\Bigl[\bm a\Bigr]\times $, which are operators that for an input vector $\bm b$ return the cross product $\bm a \times \bm b$. Because the particles are not chiral, we have $\tilde{\bm O}_{13}=0$ and there is no active torque generated by a single particle. Combining Eq.\ref{coupling_array} with Eq.\ref{coeffs}, we obtain the extended constitutive equations:
\begin{equation}\label{constitutive_N_full}
\begin{aligned}
r_i = &-\Lambda \Delta\mu_i -\Lambda' \sum_{j \in \mathcal{N}_i}\Delta\mu_j -\zeta_0 \bm p_i \cdot \bm h_i - \sum_{j \in \mathcal{N}_i} ((\zeta-\zeta_0)\bm p_j+ \zeta' \bm p_i +\zeta_r \bm r_{ji})\cdot \bm h_j -\lambda \bm p_i\cdot \bm v_i\\ 
&  +\sum_{j \in \mathcal{N}_i} \left(-\frac{\lambda'}{2}(\bm p_i-\bm p_j)+\frac{\lambda''}{2}(\bm p_i+\bm p_j)-\frac{1}{2}\lambda'_r \bm r_{ij}\right)\cdot\Delta\bm v_{ij} \\
&+\sum_{j\in \mathcal{N}_i}\left(-\frac{\mu_r}{2}\bm r_{ij}\times (\bm p_i+\bm p_j) +\frac{\mu''_r}{2}\bm r_{ij}\times (\bm p_i - \bm p_j)- \frac{\mu'_r}{2}\bm p_i\times \bm p_j\right)\cdot(\bm \Omega_i-\bm \Omega_j)~,\\
\frac{D\bm p_i}{Dt}  = & \Delta\mu_i \zeta_0 \bm p_i + \sum_{j\in \mathcal{N}_i} \Delta\mu_j((\zeta-\zeta_0)\bm p_i+ \zeta' \bm p_j + \zeta_r \bm r_{ij})+\alpha \bm h_i+\sum_{j\in\mathcal{N}_i} \alpha' \bm h_j+\chi\bm p_i \times \bm\Omega_i -\eta \bm v_i + \sum_{j\in \mathcal{N}_i} \eta'\Delta \bm v_{ij}  \nonumber\\
&+\sum_{j\in \mathcal{N}_i }(\gamma \bm p_i+\gamma' \bm p_j +\gamma'' \bm r_{ij})\times (\bm \Omega_i -\bm \Omega_j)~, \\
\bm \Gamma_i^f  = & - \chi \bm p_i \times \bm h_i-\xi_r \bm \Omega_i + \beta \bm p_i \times \bm v_i  ~, \\
\bm F_i^f  =& \Delta\mu_i  \lambda \bm p_i-\eta \bm h_i- \beta \bm p_i \times \bm \Omega_i-\xi \bm v_i    ~, \\
\bm F_{ij}^d  = & \frac{\Delta\mu_i +\Delta\mu_j}{2}\left[\lambda'(\bm p_i-\bm p_j)+\lambda'_r \bm r_{ij}\right]
-\frac{\Delta\mu_i -\Delta\mu_j}{2}\lambda''(\bm p_i+\bm p_j)+\eta' (\bm h_i -\bm h_j)
 -\xi_{||}\Delta \bm v_{ij}+\beta' (\bm p_i+\bm p_j)\times(\bm \Omega_i-\bm\Omega_j)~, \\
\bm \Gamma_{ij}^d  = & \frac{\Delta\mu_i +\Delta\mu_j}{2}\left[\mu_r\bm r_{ij}\times(\bm p_i+\bm p_j)+\mu'_r\bm p_i\times\bm p_j\right] 
- \frac{\Delta\mu_i -\Delta\mu_j}{2} \mu_r'' \bm r_{ij}\times (\bm p_i - \bm p_j)+\gamma(\bm p_j \times \bm h_j-\bm p_i \times \bm h_i)\\
& +\gamma'(\bm p_i \times \bm h_j-\bm p_j \times \bm h_i)-\gamma'' \bm r_{ij}\times (\bm h_i+\bm h_j)
-\beta' (\bm p_i+\bm p_j)\times \Delta\bm v_{ij}-\xi'_r(\bm \Omega_i - \bm \Omega_j)~.
\end{aligned}
\end{equation}
\twocolumngrid
\section{Dimensionless parameters}\label{dimensionless_appendix}
In the case of a fixed distance $2R$ between neighbouring cells, we list below the parameters of the model corresponding to the simplified constitutive equations (Eq.\ref{constitutive_N}), their interpretations and their dimensions:
\begin{itemize}
\item $R$: Cell radius ($L$).
\item $\lambda$, $\lambda'$: Active forces magnitudes ($MLT^{-2}$).
\item $\xi$, $\xi_{||}$: Translational friction coefficients ($MT^{-1}$).
\item $\xi_r$, $\xi'_r$: Rotational friction coefficient ($ML^2 T^{-1}$).
\item $\eta$, $\eta'$: Flow polarity coupling coefficient ($L^{-1}$).
\item $\zeta'$, $R\zeta_r$: Active polarity remodellings ($T^{-1}$).
\item $\alpha$: Rotational viscosity of polarity ($M^{-1}L^{-2}T$).
\item $R\mu_r$, $\mu'_r$: Active torques magnitudes ($ML^2T^{-2}$).
\end{itemize}
We discuss three different cases, for which we obtain different sets of dimensionless parameters, as discussed below. We note that as introduced in the main text, we take for these three cases $\zeta'=\xi_r'=\mu_r=\mu_r'=0$ and $\alpha\rightarrow\infty$.
\subsection{Rotation from flow alignment ($\lambda$, $\eta$)}\label{dimless-1}
In this case, we define dimensionless units in terms of $|\lambda|$, $R$, and $\xi$. We set $\lambda'$, $\eta'$, $\zeta_r$ to zero, and consider the following dimensionless parameters:
\begin{equation}
\frac{\lambda}{|\lambda|} ~,~ \frac{\xi_{||}}{\xi}~,~\frac{\xi_r}{R^2\xi}~,~\eta R~.
\end{equation}
For simplicity, we study the case where $\xi_{||}/\xi=\xi_r/(R^2\xi)=1$.
\subsection{Rotation from flow alignment ($\lambda'$, $\eta'$)}\label{dimless-2}
In this case, we define dimensionless units in terms of $|\lambda'|$, $R$, and $\xi$. We set $\lambda$, $\eta$, $\zeta_r$ to zero, and consider the following dimensionless parameters:
\begin{equation}
\frac{\lambda'}{|\lambda'|} ~,~ \frac{\xi_{||}}{\xi}~,~\frac{\xi_r}{R^2\xi}~,~\eta' R~.
\end{equation}
For simplicity, we also set $\xi_{||}/\xi=\xi_r/(R^2\xi)=1$.
\subsection{Rotation from active propulsion and polarity remodelling}\label{dimless-3}
Here, we introduce a higher-order term in the coupling coefficient for the active polarity remodelling ($\bm O_{12}^{ij}$ in Eq.\ref{coeffs}):
\begin{equation}
\bm O_{12}^{ij} = (\zeta-\zeta_0)\bm p_j+\zeta'\bm p_i+\zeta_r\bm r_{ji}+\zeta_{rp}(\bm p_j \cdot \bm r_{ji})~.
\end{equation}
This adds a parameter $\zeta_{rp}$ whose dimension is $L^{-2}T^{-1}$. We define dimensionless units in terms of $\zeta_r$ (assumed to be positive), $R$ and $\xi$. We set $\lambda'$, $\eta$, $\eta'$ to zero, leading to the following dimensionless parameters:
\begin{equation}
\frac{\lambda}{\zeta_r\xi R^2}~,~ \frac{\xi_{||}}{\xi}~,~\frac{\xi_r}{R^2\xi}~,~\frac{\zeta_{rp}R}{\zeta_r}~.
\end{equation}
For simplicity, we also set $\xi_{||}/\xi=\xi_r/(R^2\xi)=1$, apart from the case described in Fig.4c, where we discuss the effect of $\xi_{||}/\xi$.
\section{Dynamics in the plane}\label{section2D_appendix}
In this section, we show the procedure that we use to study the dynamics of a rotating doublet in the two dimensional plane. First, we show how to reduce the 3D constitutive equations to 2D equations of motions of the form of Eqs. \ref{psi1}-\ref{theta}. Then, we describe how to solve for steady-states and check their the symmetries. We find that all the steady-states have the symmetries described in Fig.2b. Finally, we discuss how to study the stability of the steady-states to small perturbations. The analytical computations were performed with Mathematica \cite{Mathematica}. The following notebooks are provided:
\begin{itemize}
\item phase\_diag\_eta.nb: Equations of motion, steady-states and linear stability in the case of section \ref{dimless-1}.
\item phase\_diag\_eta\_confined.nb: Equations of motion, steady-states and linear stability in the case of section \ref{dimless-1} with the addition of an external confinement force, imposing that the doublet center of mass is fixed.
\item phase\_diag\_etap.nb: Equations of motion, steady-states and linear stability in the case of section \ref{dimless-2}, and numerical integration of the stochastic version of the equation of motion (SM Eq.18).
\item phase\_diag\_zetarp.nb: Equations of motion, steady-states and linear stability in the case of section \ref{dimless-3}.
\item solving\_2d.nb: Generic numerical integration of the equations of motion in the plane.
\end{itemize}
\subsection{Equations of motion}
\subsubsection{3D equations of motion of a cell doublet}\label{3d_eq_appendix}
For two cells, the system's dynamics is described by the force balance and torque balance at low Reynolds number (Eqs \ref{force_bal}-\ref{torque_bal}), and the polarity dynamics equations (Eq. \ref{corot}):
\begin{equation}
\begin{aligned}
 \bm F^{f}_1+\bm F_{12} =0 &~,~ \bm  F^{f}_2-\bm F_{12} =0 ~,~\\
\bm \Gamma^f_1+\bm \Gamma_{12}=0 &~,~ \bm \Gamma^f_2+\bm \Gamma_{21}=0~,\\
\frac{d\bm p_1}{dt} = \bm \Omega_1 \times \bm p_1+\frac{D\bm p_1}{Dt}&~,~\frac{d\bm p_2}{dt} = \bm \Omega_2 \times \bm p_2+\frac{D\bm p_2}{Dt}~.
\end{aligned}
\end{equation}
The interaction forces and torques have equilibrium and deviatoric components. Combining Eqs.\ref{gamma_a}, \ref{eq_tor_for}, \ref{deviatoric}, we obtain:
\begin{equation}
\bm F_{12}=\bm F_{12}^{eq}+\bm F_{12}^d~,~\bm \Gamma_{12} = -\bm h_1^{\langle 12\rangle}\times \bm p_1 +\frac{1}{2}\bm r_{12}\times \bm F^d_{12}+\bm \Gamma_{12}^d~.
\end{equation}
The expressions of the deviatoric forces, torques, and of $D\bm p_1/Dt$, $D\bm p_2/Dt$, are given by the constitutive equations (Eq. \ref{constitutive_N}). The full dynamics in 3D is described by:
\onecolumngrid
\begin{equation}
\begin{aligned}
\lambda \bm p_1-\xi \bm v_1 -\eta \bm h_1+\bm F_{12}^{eq}+\lambda'(\bm p_1-\bm p_2)-\xi_{||}\Delta \bm v_{12}+\eta' (\bm h_1 -\bm h_2)  = 0 &~, \\
\lambda \bm p_2-\xi \bm v_2-\eta \bm h_2-\bm F_{12}^{eq}+\lambda'(\bm p_2-\bm p_1)+\xi_{||}\Delta \bm v_{12}+\eta' (\bm h_2 -\bm h_1)  = 0&~, \\
-\xi_r \bm \Omega_1-\bm h_1^{\langle 12\rangle}\times \bm p_1+\frac{1}{2}\bm r_{12}\times\left[ \lambda'(\bm p_1-\bm p_2)-\xi_{||}\Delta \bm v_{12}+\eta' (\bm h_1 -\bm h_2) \right] &  \\
+ \mu_r\bm r_{12}\times(\bm p_1+\bm p_2)+\mu'_r\bm p_1\times\bm p_2 -\xi'_r(\bm \Omega_1 - \bm \Omega_2)  = 0&~, \\
-\xi_r \bm \Omega_2-\bm h_2^{\langle 12\rangle}\times \bm p_2-\frac{1}{2}\bm r_{12}\times\left[ \lambda'(\bm p_2-\bm p_1)+\xi_{||}\Delta \bm v_{12}+\eta' (\bm h_2 -\bm h_1) \right] &  \\
- \mu_r\bm r_{12}\times(\bm p_1+\bm p_2)+\mu'_r\bm p_2\times\bm p_1 -\xi'_r(\bm \Omega_2 - \bm \Omega_1)  = 0&~, \\
\frac{d\bm p_1}{dt} = \bm \Omega_1\times \bm p_1  + \zeta' \bm p_2 + \zeta_r \bm r_{12}-\eta \bm v_1 + \eta'\Delta \bm v_{12}+\alpha \bm h_1 & ~,\\
\frac{d\bm p_2}{dt} = \bm \Omega_2\times \bm p_2 + \zeta' \bm p_1 - \zeta_r \bm r_{12}-\eta \bm v_2 - \eta'\Delta \bm v_{12}+\alpha \bm h_2 & ~.
\end{aligned}\label{3d_eq_motion}
\end{equation}
\twocolumngrid
This system of six vectorial equations can be used to determine the values of $\bm v_1$, $\bm v_2$, $\bm \Omega_1$, $\bm \Omega_2$, $d\bm p_1/dt$, $d\bm p_2/dt$, provided that an expression is given for the passive terms $\bm h_1$, $\bm h_2$, $\bm h_1^{\langle 12\rangle}$, $\bm h_2^{\langle 12\rangle}$, and $\bm F_{12}^{eq}$. Since we wish to impose $|\bm r_{12}|=2R$, $|\bm p_1|=1$ and $|\bm p_2|=1$, we introduced time dependent Lagrange multipliers $\kappa_1$, $\kappa_2$, and $\kappa_r$ in the free energy (Eq.\ref{free_energy_lagrange}). This leads to the following passive terms:
\begin{equation}
\begin{aligned}
\bm h_1 = -2\kappa_1\bm p_1~,~\bm h_2 = -2\kappa_2\bm p_2~,\\
\bm h_1^{\langle 12\rangle}=\bm h_2^{\langle 12\rangle}=0~,~\bm F_{12}^{eq} = 2\kappa_r\bm r_{12}~.
\end{aligned}
\end{equation}
In addition to $\bm v_1$, $\bm v_2$, $\bm \Omega_1$, $\bm \Omega_2$, $d\bm p_1/dt$, $d\bm p_2/dt$, we also need to solve for $\kappa_1$, $\kappa_2$, $\kappa_r$ using the following constraint equations:
\begin{equation}
\frac{d\bm p_1}{dt}\cdot \bm p_1 = 0~,~\frac{d\bm p_2}{dt}\cdot \bm p_2 = 0~,~(\bm v_1 - \bm v_2)\cdot \bm r_{12} = 0~.\label{lagrange}
\end{equation}
We note here that, for this particular choice of energy, substracting the two torque balance equations in Eq.\ref{3d_eq_motion} leads to:
\begin{equation}
\begin{aligned}
(\xi_r+2\xi'_r)(\bm \Omega_1-\bm \Omega_2) =&  2\mu_r\bm r_{12}\times(\bm p_1+\bm p_2)\\
& +2\mu'_r\bm p_1\times\bm p_2~.
\end{aligned}\label{eq_xir_noeffect}
\end{equation}
In the absence of active torques ($\mu_r=0$, $\mu'_r=0$), we have $\bm \Omega_1=\bm \Omega_2$. In this case, the dissipation term in $\xi'_r$ in the torque balance equations has no effect.
\subsubsection{Dynamics in the plane}\label{2d_dynamics_FG_appendix}
In order to study the dynamics in the $x$-$y$ plane, as shown on Fig. 2a, we introduce the angles $\theta$, $\psi_1$, $\psi_2$. Using $\bm v_g$ for the velocity of the doublet center of mass, and $\bm \Omega_d$ for the doublet angular velocity, we have:
\begin{equation}
\begin{gathered}
\bm \Omega_1 = (0,0,\Omega_1)~,~\bm \Omega_2 = (0,0,\Omega_2)~,~\bm \Omega_d=\left(0,0,\frac{d\theta}{dt}\right)~,\\
\bm v_1=\bm v_g - \frac{1}{2}\bm \Omega_d\times\bm r_{12}~,~\bm v_2=\bm v_g + \frac{1}{2}\bm \Omega_d\times\bm r_{12}~, \\
\bm p_1=(\cos\psi_1,\sin\psi_1,0)~,~\bm p_2=(\cos\psi_2,\sin\psi_2,0)~,\\
\bm r_{12} = (2R\cos \theta, 2R\sin\theta,0)~.
\end{gathered}
\end{equation}
This construction already satisfies the constraints of Eq.\ref{lagrange}. We need to use Eq.\ref{3d_eq_motion} to solve for $d\psi_1/dt$, $d\psi_2/dt$, $d\theta/dt$, $\Omega_1$, $\Omega_2$, $v_{gx}$, $v_{gy}$, $\kappa_1$, $\kappa_2$, $\kappa_r$. For convenience, we introduce short-hand notations for Eq.\ref{3d_eq_motion} in the plane:
\begin{equation}
\begin{aligned}
\bm F\bm B_1 = 0&~,~\bm F\bm B_2 = 0~,\\
TB_{1z} = 0&~,~ TB_{2z} = 0~, \\
\frac{d\bm p_1}{dt}  = \bm D\bm P_1&~,~\frac{d\bm p_2}{dt}  = \bm D\bm P_2~. \\
\end{aligned}
\end{equation}
Only the $z$ component of the torque balance equations remain. We construct the following equivalent set of equations:
\begin{equation}
\begin{gathered}
\bm F\bm B_1 + \bm F\bm B_2 = 0~,\\
(\bm F\bm B_1 - \bm F\bm B_2)\cdot \bm r_{12} = 0~,~(\bm F\bm B_1 - \bm F\bm B_2)\cdot \bm r^\perp_{12} = 0~, \\
TB_{1z} = 0~,~ TB_{2z} = 0~, \\
\bm D\bm P_1\cdot \bm p_1 = 0 ~,~ \bm D\bm P_2\cdot \bm p_2=0~,\\
\bm D\bm P_1\cdot \bm p^\perp_1 = \frac{d\psi_1}{dt} ~,~ \bm D\bm P_2\cdot \bm p^\perp_2=\frac{d\psi_2}{dt}~.
\end{gathered}
\end{equation}
we have introduced $\bm r^\perp_{12}=(-2R\sin\theta,2R\cos\theta,0)$ and $\bm p^\perp_i=(-\sin\psi_i,\cos\psi_i,0)$. This form allows to solve sequentially for the different unknowns in a relatively straightforward manner. We first solve for the center of mass motion $v_{gx}$, $v_{gy}$ using $\bm F\bm B_1 + \bm F\bm B_2 = 0$, for $\kappa_r$ using $(\bm F\bm B_1 - \bm F\bm B_2)\cdot \bm r_{12} = 0$, and for $\kappa_1$ and $\kappa_2$ using $\bm D\bm P_1\cdot \bm p_1 = 0$ and $\bm D\bm P_2\cdot \bm p_2 = 0$. We then solve for $d\theta/dt$ using $(\bm F\bm B_1 - \bm F\bm B_2)\cdot \bm r^\perp_{12} = 0$, and $\Omega_1$, $\Omega_2$ using $TB_{1z} = 0$, $TB_{2z} = 0$. The two remaining equations give the time evolution of $\psi_1$ and $\psi_2$, or equivalently $\Delta\psi_1=\psi_1-\theta$ and $\Delta\psi_2=\psi_2-\theta$ (see Fig. 2a). At this point, since $\bm h_i$ are technically Lagrange multipliers whose value is adjusted to maintain the norm of the polarities, it means that the rotational viscosity $\alpha$ does not play a role apart from renormalising the values of higher order terms in Eq.\ref{constitutive_N} that we do not consider here. We therefore take the limit $\alpha \rightarrow \infty$, for which $\bm h_i \rightarrow 0$, cancelling the terms $\eta \bm h_i$ and $\eta'(\bm h_i-\bm h_j)$.

 In the first case discussed in the main text (rotation from flow alignment, $\lambda$ and $\eta$ only, as described in section \ref{dimless-1}) we are left with the following set of differential equations:
\begin{equation}
\begin{aligned}
\frac{d\Delta\psi_1}{dt} & = A \lambda  \Bigl[2 \sin (\Delta\psi_1) \bigl[\xi  \xi_r\\
& +\eta  R \cos (\Delta\psi_2)(\xi  \xi_r+2 \xi_{||} \xi_r+2 \xi  \xi_{||}R^2)\bigr]\\
& -2 \xi_r \sin (\Delta\psi_2) (2 \eta\xi_{||} R \cos (\Delta\psi_1)+\xi ) \\
& -\eta  \xi  R \sin (2 \Delta\psi_1) (\xi_r+2 \xi_{||} R^2)\Bigr]~,\\
\frac{d\Delta\psi_2}{dt} & = B\lambda  \biggl[\sin (\Delta\psi_2) \Bigl[\eta  R \cos (\Delta\psi_1)\\
&(\xi_r (\xi +2 \xi_{||})+2 \xi  \xi_{||} R^2)-\xi \bigl[\xi_r\\
&+\eta  R \cos (\Delta\psi_2) (\xi_r+2 \xi_{||}R^2)\bigr]\Bigr]\\
&+\xi_r \sin (\Delta\psi_1) (\xi -2\eta  \xi_{||} R \cos (\Delta\psi_2))\biggr]~,\\
  \frac{d\theta}{dt}& = -C\lambda  (\sin (\Delta\psi_1)-\sin (\Delta\psi_2))~,
\end{aligned}
\end{equation}
where $A$, $B$, $C$, are functions of the friction coefficients:
\begin{equation}
\begin{aligned}
A & = \frac{1}{4 \xi  R\left(\xi_r (\xi +2 \xi_{||})+2 \xi  \xi_{||} R^2\right)}~, \\
B & = \frac{1}{2 \xi  R \left(\xi_r (\xi +2 \xi_{||})+2 \xi  \xi_{||} R^2\right)}~,  \\
C & = \frac{ \left(\xi_r+2\xi_{||} R^2\right)}{4 \xi  \xi_{||} R^3+2 \xi_r R (\xi +2\xi_{||})}~.
\end{aligned}
\end{equation}
The dynamics of $\Omega_1$, $\Omega_2$ and $\bm v_g$ are entirely determined from $\Delta\psi_1$, $\Delta\psi_2$ and $\theta$. From here, we can first make the system dimensionless (see section \ref{dimless-1}). For the sake of simplicity, we assume that the different kinds of friction are of same magnitude, and further impose that $\xi_r/(\xi R^2)=\xi_{||}/\xi=1$. We are left with the parameters $\lambda/|\lambda|$ and $\eta R$, which we simply note as $\lambda$ and $\eta$ in the following:
\begin{equation}
\begin{aligned}
\frac{d\Delta\psi_1}{d t'} & = -\frac{1}{10} \lambda  \Bigl[\sin (\Delta\psi_1) [3 \eta  \cos (\Delta\psi_1)-5 \eta \cos (\Delta\psi_2)\\
&-1]+\sin (\Delta\psi_2) (2 \eta  \cos (\Delta\psi_1)+1)\Bigr],\\
\frac{d\Delta\psi_2}{dt'} & = -\frac{1}{10} \lambda  \Bigl[\sin (\Delta\psi_1) (2 \eta  \cos (\Delta\psi_2)-1)\\
&+\sin(\Delta\psi_2) (-5 \eta  \cos (\Delta\psi_1)+3 \eta  \cos (\Delta\psi_2)+1))\Bigr],\\
   \frac{d\theta}{dt'} & = -\frac{3}{10} \lambda  (\sin (\Delta\psi_1)-\sin (\Delta\psi_2))~.
\end{aligned}\label{2d_eq}
\end{equation}
The dimensionless time $t'$ is linked to the real time $t$ by $t'=t|\lambda|/(R\xi)$. This system of equations is of the form of Eqs. \ref{psi1}-\ref{theta} with the functions $F$ and $G$ defined as:
\begin{equation}
\begin{aligned}
F(\Delta &\psi_1,\Delta \psi_2)  = -\frac{1}{10} \lambda \Bigl[\sin (\Delta \psi_1) [3 \eta  \cos (\Delta \psi_1)\\
& -5 \eta \cos (\Delta\psi_2)-1]+\sin (\Delta \psi_2) (2 \eta  \cos (\Delta \psi_1)+1)\Bigr]~, \\
G(\Delta &\psi_1,\Delta \psi_2)  = -\frac{3}{10}\lambda (\sin(\Delta \psi_1)-\sin(\Delta \psi_2))~.
\end{aligned}
\end{equation}
We also study a slightly different situation where the center of mass of the doublet is constrained to be fixed, which we associate to a situation where the doublet is confined. In this case we add an external force $\bm F_{\text{ext}}$ (exerted on both cells) in the force balance equations $\bm F\bm B_1$ and $\bm F\bm B_2$. We solve for the expression of $\bm F_{\text{ext}}$ by imposing the constraint $\bm v_g=0$. This leads to a different expression for $F$, while $G$ is unchanged:
\begin{equation}
\begin{aligned}
F & = \frac{1}{10} \lambda (1-3\eta \cos\Delta\psi_1)( \sin\Delta\psi_1-\sin\Delta\psi_2)~, \\
G& = -\frac{3}{10}\lambda (\sin(\Delta \psi_1)-\sin(\Delta \psi_2))~ .
\end{aligned}
\label{eq:case_1_confinement}
\end{equation}
For the second case discussed in the main text (rotation from flow alignment, $\lambda'$ and $\eta'$, section \ref{dimless-2}), we follow the same procedure and are left with the dimensionless parameters $\lambda'/|\lambda'|$ and $\eta'R$, which we simply note as $\lambda'$ and $\eta'$ in the following:
\begin{equation}
\begin{aligned}
F& = \frac{2}{5}\lambda' (\sin (\Delta \psi_1)-\sin (\Delta \psi_2)) (2 \eta' \cos (\Delta\psi_1)+1)~, \\
G & = -\frac{1}{5} \lambda' (\sin(\Delta \psi_1) - \sin(\Delta \psi_2))~,
\end{aligned}\label{eq_lambdap}
\end{equation}
with a different time normalisation $t'=t|\lambda'|/(R\xi)$.
Finally, in the case of self-propulsion and active polarity remodelling (see section \ref{dimless-3}), we are left with the dimensionless parameters $\lambda/(\zeta_r \xi R^2)$ and $\zeta_{rp}R/\zeta_r$, which we simply note $\lambda$ and $\zeta_{rp}$ in the following:
\begin{equation}
\begin{aligned}
F & = \frac{1}{10} [(-20 + \lambda) \sin(\Delta \psi_1) - 20 \zeta_{rp} \sin(2\Delta\psi_1) \\
&- \lambda \sin(\Delta \psi_2)]~, \\
G& = -\frac{3}{10}\lambda (\sin(\Delta \psi_1)-\sin(\Delta \psi_2))~,
\end{aligned}\label{eq_last_case}
\end{equation}
with time normalisation $t'=t \zeta_r R$.
\subsection{Steady-states and their symmetries}\label{ss-sym}
We continue to use the example of the first case discussed in the main text (Eq.\ref{2d_eq}) to explain how we look for steady-states. We consider any configuration having $d\Delta\psi_1/dt=d\Delta\psi_2/dt=0$ as a steady-state even if $d\theta/dt\neq 0$, so that we can account for rotating steady-states. We introduce the variables $c_1=\cos(\Delta\psi_1)$, $s_1=\sin(\Delta\psi_1)$, $c_2=\cos(\Delta\psi_2)$, $s_2=\sin(\Delta\psi_2)$. A configuration characterised by $c_1$, $s_1$, $c_2$, $s_2$ is a steady-state if and only if:
\begin{equation}
\begin{aligned}
-\frac{1}{10} \lambda  (s_1 (3 c_1 \eta -5 c_2 \eta -1)+s_2 (2c_1\eta +1))=0~,\\
-\frac{1}{10} \lambda  (s_2 (-5 c_1 \eta +3 c_2 \eta +1)+s_1 (2c_2 \eta -1))=0~,\\
c_1^2+s_1^2=1~,~c_2^2+s_2^2=1~,~(c_1,s_1,c_2,s_2)\in \mathbb{R}^4~.
\end{aligned}
\end{equation}
When solving analytically for the all the possible solutions (allowing non-real ones), we find twenty four distinct solutions. We first check whether there exists solutions that do not satisfy any symmetry shown on Fig. 2b (``rotating", ``flocking", ``mirror"). The criteria on $c_1$, $s_1$, $c_2$, $s_2$ for each symmetry are:
\begin{equation}
\begin{aligned}
\text{``rotating"} &\Leftrightarrow c_1=-c_2 &\text{ and }&s_1=-s_2~,&s_1\neq 0,\\
\text{``flocking"} &\Leftrightarrow c_1=c_2 &\text{ and }&s_1=s_2~,&\\
\text{``mirror"} &\Leftrightarrow c_1=-c_2 &\text{ and }&s_1=s_2~.&\\
\end{aligned}
\end{equation}
From this we infer that the criteria for having none of these symmetries is:
\begin{equation}
\begin{aligned}
\text{``no symmetry"}\Leftrightarrow & (c_1 + c_2 \neq 0 \text{ and } (c_1 \neq c_2 \text{ or } s_1 \neq s_2))\\
 \text{ or }& (s_1 + s_2 \neq 0 \text{ and } s_1 \neq s_2)~.
 \end{aligned}
\end{equation}
Out of the twenty four solutions, we verified that four of them satisfy the ``no symmetry" criterion. We then checked whether these four remaining solutions could verify $(c_1,s_1,c_2,s_2)\in \mathbb{R}^4$. We found that they did not. This showed that all the steady-states are of the type ``rotating", ``flocking" or ``mirror". This was found to be true not only in the context of Eq.\ref{2d_eq} but also in the other cases discussed in the main text. We refer to Mathematica notebooks for further details on these results.
\subsection{Linear stability analysis}
We study the stability of the steady-states of the type ``rotating", ``mirror" and ``flocking", for each of the cases discussed in the main text. This procedure allows to compute the regions shown in the phase diagrams of Figs. \ref{phasediagetap}, \ref{phasediag_zetarp}. As an example, we focus again on the case of Eq.\ref{2d_eq}. We only consider the dynamics of $\Delta\psi_1$ and $\Delta\psi_2$. Assuming that we are in a ``rotating" steady-state, we have $\Delta\psi_2=\Delta\psi_1+\pi$. This leads to the equation for the steady-state angle $\Delta\psi_1$:
\begin{equation}
\frac{1}{5} \lambda  \sin(\Delta\psi_1) (1-3 \eta  \cos (\Delta\psi_1))=0~.
\end{equation}
The trivial solutions $(\Delta\psi_1,\Delta\psi_2)=(0,\pi)$, $(\Delta\psi_1,\Delta\psi_2)=(\pi,0)$, which are special cases which we name ``in" and ``out", must be studied separately from the other solution which is:
\begin{equation}
\Delta \psi_1^\star = \cos^{-1}\left(\frac{1}{3\eta}\right)~,
\end{equation}
The equivalent solution $\Delta\psi_1=-\Delta\psi_1^\star$ only changes the sign of $d\theta/dt$ and is not discussed further. We now introduce small displacements $\epsilon_1$, $\epsilon_2$ around the steady-state such that:
\begin{equation}
\Delta\psi_1  = \Delta\psi_1^\star + \epsilon_1~,~ \Delta\psi_2 = \pi+\Delta\psi_1^\star+\epsilon_2~.
\end{equation}
Expanding Eqs. \ref{psi1}-\ref{theta} around the steady-state, we have:
\begin{equation}
\begin{aligned}
\frac{d\epsilon_1}{dt} & = F(\Delta\psi_1^\star+\epsilon_1,\pi+\Delta\psi_1^\star+\epsilon_2)~,\\
\frac{d\epsilon_2}{dt} & = -F(-\Delta\psi_1^\star-\epsilon_2,\pi-\Delta\psi_1^\star-\epsilon_1)~.
\end{aligned}\label{reduced_2d_eq}
\end{equation}
After linearisation to first order in $\epsilon_1$, $\epsilon_2$, in matrix notation we obtain:
\begin{equation}
\begin{pmatrix}
\frac{d\epsilon_1}{dt}\\
\frac{d\epsilon_2}{dt}
\end{pmatrix}=M
\begin{pmatrix}
\epsilon_1\\
\epsilon_2
\end{pmatrix}~,
\end{equation}
where $M$ is the following matrix:
\begin{equation}
\begin{pmatrix}
\partial_1F(\Delta\psi_1^\star,\pi+\Delta\psi_1^\star) & \partial_2F(\Delta\psi_1^\star,\pi+\Delta\psi_1^\star)\\
\partial_2F(-\Delta\psi_1^\star,\pi-\Delta\psi_1^\star) & \partial_1F(-\Delta\psi_1^\star,\pi-\Delta\psi_1^\star)
\end{pmatrix},
\end{equation}
where $\partial_i F$ is the partial derivative of $F$ with respect to the $i$th argument. A steady-state is stable if and only if the eigenvalues of the matrix above have negative real parts. In the case of Eq.\ref{reduced_2d_eq}, around $\Delta\psi_1^\star$, the eigenvalues are:
\begin{equation}
-\frac{\left(6 \eta ^3+\eta \right) \lambda }{15 \eta ^2},\frac{\left(9 \eta ^3-\eta \right) \lambda }{15 \eta ^2}
\end{equation}
The eigenvalues are always real, but for them to be both strictly negative, $\eta$ and $\lambda$ need to verify:
\begin{gather}
\left(\lambda <0\text{ and } -\frac{1}{3}< \eta <0\right)\\
\text{ or } (\lambda =0\text{ and } (\eta <0 \text{ or }
   \eta >0)) \\
   \text{ or } \left(\lambda >0 \text{ and } 0<\eta < \frac{1}{3}\right)~.
\end{gather}
In addition, the solution $\Delta\psi_1^\star$ must be real, which requires $\eta \leq -1/3$ or $\eta \geq 1/3$. We therefore see in that case that there is never a stable rotating state.

In the case where the center of mass of the doublet does not move, the rotating solutions with $\Delta \psi_1=\Delta \psi_1^\star $, $\Delta \psi_2=\pi+\Delta \psi_1^\star $ and $\Delta \psi_1=-\Delta \psi_1^\star$, $\Delta \psi_2=\pi-\Delta \psi_1^\star $ are still steady-states, with stability now determined by a single eigenvalue:
\begin{align}
\frac{\left(9 \eta ^2-1\right) \lambda }{15 \eta }
\end{align}
such that the rotating solution exists and is stable for $\lambda<0, \eta>1/3$ or $\lambda>0$, $\eta<-1/3$.
\bibliography{main}
\end{document}